\documentclass[aps,pra,twocolumn,showpacs,amsmath,amsfonts,amssymb,floatfix,superscriptaddress,nofootinbib]{revtex4}

\usepackage{bm}																
\usepackage{color}
\usepackage{float}
\usepackage{graphicx} \graphicspath{{./Images/}}
\usepackage{hyperref}
\usepackage[capitalise]{cleveref}

\usepackage{tikz} \usetikzlibrary{arrows}
\usepackage{etoolbox}														
\preto\tikzpicture{\catcode`$=3 }												
\preto\tikz{\catcode`$=3 }														

\begin{document}

%--------------------------------------------------------------------------------------------------------------------------------------------------%

\newcommand \br {\mathbf{r}}

\newcommand{\sumN}{\sum\limits_{i=1}^N }

\renewcommand\Im{\operatorname{Im}}

\def \bra<#1|{\langle #1 \vert }
\def \ket|#1>{\vert #1 \rangle }
%\NewDocumentCommand \braket{r<>}{\langle \StrSubstitute{#1}{|}{\vert } \rangle}
%\RenewDocumentCommand \outer{r|>r<|}{\vert #1 \rangle \langle #2 \vert}
\def \Bra<#1|{\left\langle #1 \right\vert }
\def \Ket|#1>{\left\vert #1 \right\rangle }
%\NewDocumentCommand \Braket{r<>}{\left\langle \StrSubstitute{#1}{|}{\middle\vert } \right\rangle}
%\NewDocumentCommand \Outer{r|>r<|}{\left\vert #1 \middle\rangle \middle\langle #2 \right\vert}

\newcommand \blue[1]{{\color{blue} #1}}
\newcommand \red[1]{{\color{red} #1}}

%--------------------------------------------------------------------------------------------------------------------------------------------------%
\title{Quantum Control of Many-Body Systems by the Density}   
\author{S. E. B. Nielsen}																	
\affiliation{Institut f\"ur Theoretische Physik, Universit\"at Innsbruck - Technickerstra{\ss}e 25, 6020 Innsbruck, Austria}
\author{M. Ruggenthaler}
\affiliation{Institut f\"ur Theoretische Physik, Universit\"at Innsbruck - Technickerstra{\ss}e 25, 6020 Innsbruck, Austria}
\author{R. van Leeuwen}
\affiliation{Department of Physics, Nanoscience Center, University of Jyv\"askyl\"a, 40014 Jyv\"askyl\"a, Finland}
\affiliation{European Theoretical Spectroscopy Facility (ETSF)}

\date{\today}

\begin{abstract}						
In this work we focus on a recently 
introduced method \cite{Nielsen2012} to construct the external potential 
$v$ that, for a given initial state, produces a prescribed 
time-dependent density in an interacting quantum many-body system. We show how
this method can also be used to perform flexible and efficient quantum control.
The simple interpretation of the density (the amount of electrons per volume) 
allows us to use our physical intuition to consider interesting control problems
and to easily restrict the search space in optimization problems. 
The method's origin in  time-dependent density-functional theory makes studies 
of large systems possible. We further discuss the generalization of the method 
to higher dimensions and its numerical implementation in great detail. 
We also present several examples to illustrate the flexibility, and to confirm that the scheme 
is efficient and stable even for large and rapid density variations irrespective 
of the initial state and interactions. 
\end{abstract}
\maketitle

%--------------------------------------------------------------------------------------------------------------------------------------------------%
\section{Introduction}
%--------------------------------------------------------------------------------------------------------------------------------------------------%

Our capability to perform accurate computer simulations of physics at the atomic scale is crucial for modern technology			
as such simulations have the potential to replace expensive laboratory tests. Ideally, one could for example investigate
all the processes that take place in a computer chip or when a drug is administered by simply running a computer program.
Such simulations are however strongly limited by the fact that most quantum mechanical problems are time-dependent
many-body problems, with wave-function formulations that quickly grow computationally intractable with system size.
Finding the physical properties of large systems thus remains a bottleneck for further progress in many areas of physics,
chemistry and many derived fields, motivating alternative approaches like time-dependent density-functional theory (TDDFT)
\cite{CarstenBook, TDDFT}. This theory allows for cheaper many-body calculations, albeit with less control over the precision 
as it relies on approximations to the so-called exchange-correlation potential.

Based on a recent fixed-point formulation of TDDFT \cite{GFPP, GFPP1D},
we presented a numerical method in \cite{Nielsen2012} to construct the external potential that,
for a given initial state, produces a prescribed time-dependent density in an interacting quantum many-body system.
This novel procedure is highly efficient and stable even for large and rapid density variations irrespective 
of the initial state and two-electron interaction. It allows us to efficiently compute any property as a functional of the 
initial state and density, with almost arbitrary accuracy if desired, including the exact Kohn-Sham and exchange-correlation 
potentials. This is an explicit realisation of the Runge-Gross result \cite{RungeGross} that any observable is a functional 
of the density and initial state, and allows us to investigate virtually all aspects of TDDFT in an unprecedented way, in the end 
hopefully leading to better, more physically motivated, exchange-correlations potentials (that it can benchmark).

%--------------------------------------------------------------------------------------------------------------------------------------------------%
								
Now, to design a computer chip or drug, it is not only important to be able to simulate what happens						
when we apply a certain external potential to a quantum system (as is the usual application of TDDFT).
Often, we need to answer the reverse question, known as control theory. Namely, if we want a system to react in a 
certain way, like going from its ground state to a specific excited state, what external potential should we apply to it?

While our numerical method has its origin in TDDFT, and TDDFT was the main focus of our original letter \cite{Nielsen2012},
it also represents a very interesting novel control scheme, which will be the main focus of the present paper.
Indeed, as the density forms a very convenient control objective, it strongly augments the many important 
control schemes \cite{OCT1, OCT2, LCT1,LCT2} already of extensive use in chemistry,
laser physics, quantum optics \cite{Mancini2005} and the physics of ultracold gases \cite{CHU02}.					
Since the density has a  particularly simple interpretation (the amount of electrons per volume),  
it is one of the most intuitive observables to prescribe what we want a system to do.
For instance, if we want only a specific part of a molecule to rotate, this is easily described by the density.
Further, in our procedure we prescribe a specific path, i.e. we prescribe what the density must be at all times.
This is called tracking, and is known for its great efficiency, as it can be implemented locally, i.e. stepwise in time.
Hence it is also known as a local control theory (LCT). Now, the crucial reason that tracking is not more used, in spite of its 
great efficiency and conceptual simplicity, is that standard tracking schemes lack the crucial guarantee that their 
control objectives actually can be achieved, leading to failure or at best major numerical challenges.
There is simply no potential that can achieve the desired path, unless the path is prescribed with appropriate care.
The density is thus special in that the van Leeuwen theorem of TDDFT \cite{vanLeeuwen} in practise always guarantees 
the existence of such a potential for this tracking objective, making tracking possible (in TDDFT terms the density must 
be $v$-representable, which is a very mild restriction). In control theory (unlike TDDFT), we are also often more interested 
in an optimal rather than a prescribed path. For instance, we may ask for the least energetic external potential that brings 
a system from its ground state to a specific excited state or transfers charge from one part of a molecule to another 
(achieved by following some optimal path).	 The standard optimal control theory (OCT) methods, however, require thousands   
of global iterations \cite{OCT1}, and are thus only feasible for very small systems. Therefore a physically reasonable restriction 
of the search space is desirable.  While such a restriction in terms of the wave functions is very challenging, 
the restriction to certain density profiles is much simpler. Since we know the set of densities that can be achieved
we can use tracking to find the external potential that yields these densities, thereby essentially doing OCT in an 
efficient, local, manner.	Thus we gain both the efficiency of LCT and flexibility of OCT (to a very high extend).

We point out that our method goes beyond the usual dipole approximation. 
While the dipole approximation is a very important special case, as it is generally the dominant laser-molecule interaction,
treating more advanced fields is fundamentally important, as a tool to gain further insight into quantum dynamics and 
to allow for more general local-control objectives. That is, given a dipole field $\bm{E}(t)$, only a quantity like 
the dipole moment $\bm{\mu}(t)$ can be locally controlled, while given a general potential $v(\br t)$, a quantity 
like the density $n(\br t)$ can be controlled (the freedom of the control field and objective must match). Such control
objectives become experimentally important, for instance, in the physics of ultracold gases, where a high temporal and spatial
control over the applied external fields is possible.
In the dipole case, our approach reduces to a known method to control the dipole moment.					                   
However, by connecting it with TDDFT, some less well-known crucial properties of that approach become immediately clear, 
e.g., that dipole tracking is one of the few other cases besides the density that are controllable.

Finally, to make the control of realistic systems feasible,
our method should be compatible with an efficient many-body method.
But, since the present method was derived within the framework of TDDFT,
it can trivially take advantage of any TDDFT approximation,
and thus makes ab-initio control of big quantum systems possible.

%--------------------------------------------------------------------------------------------------------------------------------------------------%
																					
\textbf{Outline} - In \cref{sec:LCT} we review LCT in its simplest form to establish its basic idea.			
In \cref{sec:TP} we then advance this idea by presenting different more advanced strategies to implement LCT,				
since a naive implementation in our case turns out to be extremely unstable. Based on this, we then present the
final algorithm in \cref{sec:DLCT}. Then, in \cref{sec:TDDFT}, 
we show how we can work with an auxiliary non-interacting system instead of the true interacting system 
using TDDFT approximations, before we in \cref{sec:OCT} present different strategies to do OCT with 
our LCT. Finally we give examples for the different possible applications of our LCT in \cref{sec:Examples}. 
In the appendices we provide details about boundary conditions, the numerical spatial representation
and different propagation routines.

%--------------------------------------------------------------------------------------------------------------------------------------------------%
\section{Local Control Theory} \label{sec:LCT}					
%--------------------------------------------------------------------------------------------------------------------------------------------------%

Throughout this paper we study the class of $N$-electron systems governed by a time-dependent 
non-relativistic spin-free Hamiltonian (atomic units are used throughout)
\begin{equation} \label{Hamiltonian}
\hat H(t) = \hat T + \hat W + \hat V(t) = - \frac{1}{2}\sumN \nabla_i^2 + \sum\limits_{i>j=1}^N w(r_{ij}) + \sumN v(\br_i t),
\end{equation}
where $\hat T$ is the kinetic energy operator,  $\hat W$ the two-electron interaction operator, 
$\hat V(t)$ the external potential operator, $\nabla_i$ the gradient with respect to the spatial coordinate 
$\br_i$ and $r_{ij} = |\br_i - \br_j|$. We usually take the internal two-electron interaction to 
be purely Coulombic, i.e. $w(r_{ij}) = r_{ij}^{-1}$, but other options are also possible, of which zero interaction 
turns out to be most important.

To review tracking in its simplest form, we will for now further assume											
the only freedom we have to control the dynamics of the quantum system is to adjust the field strength 
$\varepsilon(t)$ of a linearly polarised laser field (say in the $x$-direction) in dipole approximation, i.e.,
\begin{equation} \label{DipoleV}
\hat V(t) = \hat V_0(t) - \varepsilon(t) \hat \mu_x ,
\end{equation}
where $\hat V_0(t)$ is fixed and usually time-independent.									
Here $\hat\mu_x  = - \sum_{i=1}^N x_i$ is the $x$-component of the dipole operator.
With $\hat H_0(t) = \hat T + \hat V_0(t) + \hat W$, the Hamiltonian then reads
\begin{equation}
\label{DipoleHamiltonian}
\hat H(t) = \hat H_0(t) - \varepsilon(t) \hat\mu_x .
\end{equation}	
This is not a lot of freedom, and hence it is also quite limited which control objectives can actually be achieved.				
The control objective must be a scalar as it cannot be more general than the freedom we 
have to achieve it.

Accordingly, we consider the control objective of obtaining a prescribed time-evolution
of the expectation value $S(t) = \bra<\Psi(t)|\hat O(t) \ket|\Psi(t)>$ of a scalar observable $\hat O(t)$ 
\footnote{$S(t)$ must of course be consistent with the initial state $\ket|\Psi_0>$,
i.e., $S(t_0) = \bra<\Psi_0|\hat O(t_0) \ket|\Psi_0>$ where $t_0$ is the initial time.}.   
This observable $\hat O(t)$ could for example be the projection operator onto the first excited state $\ket|1>\bra<1|$,	
in which case $S(t)$ is simply the population of the first excited state.
If we then prescribe that $S(t)$ should go from $0$ to $1$ in some specific way, and we start out in the ground state,
well then we are simply asking that the system goes from its ground state to first excited state in this specific manner.

To actually obtain the field strength $\varepsilon(t)$ that achieves this control objective, there exist a clever combination of tricks.
We start by taking the time-derivative of $S(t)$, using the time-dependent Schr\"odinger equation (TDSE),
\begin{equation*}						
i \partial_t \Ket|\Psi(t)> = \left[\hat H_0(t) - \varepsilon(t) \hat\mu_x \right] \Ket|\Psi(t)> ,
\end{equation*}
for the time-derivatives of the wave function. This yields the Ehrenfest theorem for $S(t)$,
\begin{equation} \label{Ehrenfest}
\partial_t S = \bra<\Psi|\partial _t \hat O - i\left[\hat O,\hat H_0 \right] + i\varepsilon \left[\hat O,\hat \mu_x \right] \ket|\Psi> ,
\end{equation}
where we can easily isolate the field strength $\varepsilon(t)$
\begin{equation} \label{DipoleTracking}
\varepsilon(t) = \frac{\partial_t S(t) + \bra<\Psi(t)| i \left[\hat O(t),\hat H_0(t) \right] - \partial _t \hat O(t) \ket|\Psi(t)>}{\bra<\Psi(t)|i\left[\hat O(t),\hat \mu_x \right] \ket|\Psi(t)>} .
\end{equation}
Hence we obtain an equation for the desired field strength $\varepsilon(t)$ in terms of only the known $S(t)$ and wave function $\ket|\Psi(t)>$.
So our only problem in calculating the desired field strength $\varepsilon(t)$ is that we do not know the wave function $\ket|\Psi(t)>$
that corresponds to this field-strength. However, numerically this problem essentially resolves itself when using time-stepping, as we shall now see.

To keep things as simple as possible for now, let us here consider Euler time-stepping on an equidistant time-grid with points $t_n = n \Delta t$.
We start by calculating $\varepsilon(t_0)$  at the initial time $t=t_0$ from the initial state $\ket|\Psi(t_0)>$ by \cref{DipoleTracking},
and then we use Euler time-stepping with the corresponding Hamiltonian $\hat H(t_0)$ to find $\ket|\Psi(t_1)>$ as
$\ket|\Psi(t_1)> = \ket|\Psi(t_0)> - i \Delta t \hat H(t_0) \ket|\Psi(t_0)>$.
By repeating this, we can simply step through time, obtaining the desired field strength $\varepsilon(t)$ as we go.

So the tracking idea is both conceptually very simple and computationally efficient (we simply step through time once and we are done).
However, there is an issue. Even if we have ensured that the freedom of the control objective and control field match,
there is still generally no guarantee that the control objective is achievable,
since $\varepsilon(t)$ can become infinite, when \cref{DipoleTracking} is singular.
Our population control example thus generally
 \footnote{A way to circumvent this problem in the case of a time-independent operator 
$\hat O$ commuting with $H_0$, like the operator $\ket|1>\bra<1|$, with a similar 
efficiency as tracking is to use $\varepsilon(t)  =  - i\bra<\Psi(t)| [\hat O,\hat \mu_x] \ket|\Psi(t)>$. 
By \cref{Ehrenfest} this ensures $\partial_t S(t) \ge 0$, i.e., $S(t)$ is monotonically increasing. 
This means that the population always gets closer to its maximal value $1$. 
This does however leave the exact form of the path open.
Note that in general the assumptions on $\hat O(t)$ are needed for this idea,						
as this maximization of the expectation value only has sense for time-independent operators,
and no field strength can make $\dot S(t) \ge 0$ if $i \bra<\Psi(t)|[\hat O,\hat \mu_x] \ket|\Psi(t)> = 0$ and $-i \bra<\Psi(t)|[\hat O,\hat H_0] \ket|\Psi> < 0$.
The reason for using this alternative expression for $\varepsilon(t)$					
is that it unlike \cref{DipoleTracking} is singularity-free.} fails \cite{LCT2}!

On the other hand, the dipole moment $\mu_x(t)$ and the density $n(\br t)$ are very special since they couple directly to the
external control fields $\varepsilon(t)$ and $v(\br t)$ in \cref{DipoleHamiltonian} and \cref{Hamiltonian} respectively. 
These so-called conjugate observables $\mu_x(t)$ and $n(\br t)$ can be shown to be practically almost always achievable
by the  the van Leeuwen theorem of TDDFT \cite{vanLeeuwen, ruggenthaler2010general}. This special property is based on the fact that two different external fields 
of a specified form lead usually to different conjugate observables\footnote{Other examples are the restriction to 
external fields of the form $\varepsilon(t) x^k$ which then have as possible conjugate observable $-\int x^k n(\br t) d\br$.}.
This is nicely reflected in non-singular tracking equations (analogues of \cref{DipoleTracking}),
which means there always exists a finite field strength giving the desired change in $S(t)$. Hence the control objective 
is indeed possible. 
To see this in the dipole case, first note that for $\hat O(t) = \hat \mu_x$ and hence $S(t)=\mu_x(t)$ the Ehrenfest theorem becomes		
\begin{equation*}																						
\partial_t \mu_x = - i \bra<\Psi|\left[\hat \mu_x,\hat T \right] \ket|\Psi> = - p_x .
\end{equation*}
So in this case we take the second time-derivative of $\mu_x(t)$ by applying the Ehrenfest theorem to $-p_x$, to obtain
\begin{equation*}
\partial_t^2 \mu_x = \bra<\Psi| i\left[\hat p_x,\hat V_0 + \hat W \right] - i\varepsilon \left[\hat p_x,\hat \mu_x \right] \ket|\Psi>	 .
\end{equation*}
The term $\hat W$ vanishes since it can be expressed as a divergence which integrates to zero \cite{tokatly2005quantum}.
By then introducing the one-electron density operator 									
\begin{equation}
\hat n(\br) = \sumN \delta (\br - \br_i) ,
\end{equation}
we easily find the promised singularity free equation for the desired tracking field strength \cite{ruggenthaler2010general}
\begin{equation} \label{DipoleTrackingDipole}
\varepsilon(t) = \frac{1}{N}\partial_t^2 \mu_x(t) - \frac{1}{N} \int n(\br t) \partial_x v_0(\br t) d\br .
\end{equation}

To see how we can derive an analogue of \cref{DipoleTracking} in the density case,	
and that it indeed also is singularity free, the story goes exactly the same as in the dipole case.
We will simply take the second time-derivative of $n(\br t)$, exactly as we did of $\mu_x(t)$ to obtain 
\cref{DipoleTrackingDipole}\footnote{Indeed, as $n(\br t)$ and $\mu_x(t)$ are related 
by $\mu_x(t) = - \int d\br x n(\br t)$, so are \cref{DipoleTrackingDipole,nvq}. 
To see this explicitly, multiply \cref{nvq} by $x$ and integrate (using partial integration for the two first terms),
\begin{equation*}
\int n(\br t) \partial_x v(\br t) d\br = \int Q_x([v],\br t)d\br + \partial_t^2 \mu_x(t) .
\end{equation*}	
The $Q_x$-term vanish since it also can be written as a divergence, so this is indeed a 
rearrangement of \cref{DipoleTrackingDipole},
when we in accordance with \cref{DipoleV} set $v(\br t) = v_0(\br t) + \varepsilon(t) x$.
In this sense, dipole tracking may indeed be viewed as a special case of density tracking.}.
To do this, we first introduce the current operator
\begin{equation*}
\hat{\mathbf{j}} (\br) =
\frac{1}{2i} \sum_{l=1}^{N} \left( \delta(\br-\br_l) \overrightarrow{\nabla}_l -
\overleftarrow{\nabla}_l \delta(\br-\br_l) \right) .
\end{equation*}
The expectation values $n(\br t)$ and $\mathbf{j}(\br t)$ of the density and current operators
satisfy equations of motion given by
\begin{align}
\label{cont1}
\partial_t n(\br t) &= -\nabla \cdot \mathbf{j} (\br t) \\
\label{cont2}
\partial_t \mathbf{j} (\br t) &= - n(\br t)\nabla v (\br t) + \mathbf{Q} (\br t) ,
\end{align}
where $\mathbf{Q}(\br t)$ is the expectation value of the internal local-force operator
$\hat{\mathbf{Q}}(\br) = -i [\hat{\mathbf{j}} (\br), \hat T + \hat W]$.
\cref{cont1,cont2} imply
\begin{equation} \label{nvq}
-\nabla \cdot \left( n(\br t)\nabla v(\br t) \right) = q([v] , \br t) - \partial_t^2 n(\br t),
\end{equation}
where $q([v],\br t) = -\nabla \cdot \mathbf{Q} (\br t)$.
For a fixed density and initial state this is an implicit equation for the potential $v$
when we regard $q([v],\br t)$ as a functional of $v$ through the time-dependent many-body state $| \Psi ([v],t) \rangle$
obtained by the TDSE with potential $v$ and given fixed initial state. Indeed, the operator 
$\nabla \cdot  n(\br t)\nabla$ can be inverted (see also \cref{sec:BC})        
and thus \cref{nvq}	 is singularity free.

So in principle we can now already do density tracking. There is but one issue;
a naive implementation like this, using Euler time-stepping, turns out to be very unstable in the density case (unlike in the other cases presented).
While the Euler method mathematically is guaranteed to converge for small enough time-steps
(at least when we stay away from singularities), it is numerically very inefficient and can fail in practice due to round-off errors,	
which it does in the density case. This comes back to the fact that the density (at a given point) may change by orders of magnitude,						
so we have to be very precise for it to stay correct.  If we do not, an extremely strong artificial potential is needed to compensate for it.
So, in order to stabilize the scheme in the density case, and significantly increase the performance in all the cases,
we discuss various advanced time-stepping strategies in the next \namecref{sec:TP}.
We then return to how the density case is really done in \cref{sec:DLCT}.
As a convenient extra bonus, the change of time-step strategy also allows us to get rid of the complicated quantity $q$.

%--------------------------------------------------------------------------------------------------------------------------------------------------%
\section{Time-Step Procedures} \label{sec:TP}					% Time-Stepping Strategies
%--------------------------------------------------------------------------------------------------------------------------------------------------%

\begin{figure}							
\begin{tikzpicture}[scale=1.5]
\node (A) at (0,3) {$A)$};
	\node (A0) at (0.5,3) {$\Psi$};
	\node (A1) at (1.5,3) {$\Psi$};
	\node (A2) at (2.5,3) {$\Psi$};
	\node (A3) at (3.5,3) {$\Psi$};
		\path[->,font=\scriptsize,>=angle 90]
		(A0) edge node[above]{$\bar{H}$} (A1)
		(A1) edge node[above]{$\bar{H}$} (A2)
		(A2) edge node[above]{$\bar{H}$} (A3);
\node (B) at (0,2) {$B)$};
	\node (B0) at (0.5,2) {$\Psi$};
	\node (B1) at (1.5,2) {$\Psi$};
	\node (B2) at (2.5,2) {$\Psi$};
	\node (B3) at (3.5,2) {$\Psi$};
		\path[->,font=\scriptsize,>=angle 90]
		(B0) edge node[above left] {$\hat{H}$} node[below] {$(\bar{H})$} (B1)
		(B1) edge node[above left] {$\hat{H}$} node[below] {$(\bar{H})$} (B2)	
		(B2) edge node[above left] {$\hat{H}$} node[below] {$(\bar{H})$} (B3);
\node (C) at (0,1) {$C)$};
	\node (CI) at (1.0,0.6) {$\Psi_0$};		
	\node (C0) at (0.5,1) {$\Psi$};
	\node (C1) at (1.5,1) {$\Psi$};
	\node (C2) at (2.5,1) {$\Psi$};
	\node (C3) at (3.5,1) {$\Psi$};
		\path[->,font=\scriptsize,>=angle 90]
		(CI) edge (C0)
		(CI) edge (C1)
		(C0) edge [bend left] node[above] {$\hat{H}$} (C2)
		(C1) edge [bend left] node[above] {$\hat{H}$} (C3);
\node (D) at (0,0) {$D)$};
	\node (D0) at (0.5,0) {$\Psi$};
	\node (D1) at (1.5,0) {$\Psi$};
	\node (D2) at (2.5,0) {$\Psi$};
	\node (D3) at (3.5,0) {$\Psi$};
		\path[->,font=\scriptsize,>=angle 90]
		(D0) edge node[above]{$\bar{H}_k$} (D1)
		(D1) edge node[above]{$\bar{H}_k$} (D2)
		(D2) edge node[above]{$\bar{H}_k$} (D3);

\draw[->,font=\scriptsize,>=angle 90] (0.5,-0.5) node[below]{$0$} -- (1.5,-0.5) node[below]{$\Delta t$} -- (2.5,-0.5) node[below]{$2\Delta t$} -- (3.5,-0.5) node[below]{$3\Delta t$} -- (4.0,-0.5) node[right]{$t$};

\foreach \x in {0.5,1.5,2.5,3.5}
	\draw (\x,-0.55) -- (\x,-0.45);
\end{tikzpicture}
\caption{Different time-stepping strategies. $\hat H$ are on-point Hamiltonians and $\bar H$ are midpoint Hamiltonians respectively.
A) Standard time-stepping strategy for known Hamiltonians.  B) Using the on-point Hamiltonian for
making a time step and then use this state to prescribe an approximate midpoint Hamiltonian. 
C) Extrapolating a new initial state between $0$ and $\Delta t$ and using
the on-point Hamiltonians as midpoint Hamiltonians for two parallel time-stepping procedures. D) Making an initial guess for the midpoint
Hamiltonian and updating it until the control target at the next time step is achieved.}									
\label{fig:strategy}
\end{figure}
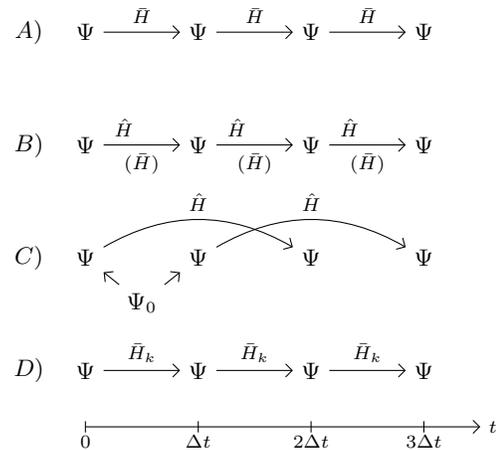

Formally, the exact time-evolution operator is given by
\begin{equation}
\hat{U}(t_0,t) = \mathcal{T} \exp \left(-i\int_{t_0}^t dt' \hat{H}(t') \right) ,
\end{equation}
where time-ordering $\mathcal{T}$ is needed unless $[\hat{H}(t),\hat{H}(t')] = 0$.		
But how do we apply this operator in a simple way?
Well, let us first of all (like previously in the Euler case) split the time-axis into equidistant pieces of length $\Delta t$,
$t_n=n\Delta t$, as illustrated in \cref{fig:strategy}\footnote{All strategies presented here can easily use adaptive time-steps 
to significantly improve the performance, except the one of \cref{fig:strategy}~C.}.
It can then be shown that for $\Delta t \to 0$ the following procedure yields the exact time-propagation\footnote{A prerequisite 
that this is true is that the Hamiltonian does not change discontinuously in time, i.e., the potential
$v(\br t)$ is at least continuous in time \cite{reed1975methods}. In practise $\Delta t$ should be so small that $\hat H(t)$ almost stays constant 
within each piece.}:

Given the initial state $\ket|\Psi_0>$ at time $t_0$, we can simply propagate the wave function to time $t_1$ by using 
the time-evolution operator for a time $\Delta t$, $\hat U(\Delta t) = e^{ - i \hat H \Delta t}$, which we from now on 
will call the time-step operator\footnote{The exact time-evolution operator for small enough $\Delta t$ may also be written as $\exp(\Omega)$,
with a convergent infinite series for $\Omega$, known as a Magnus series. Truncating this series to 2\textsuperscript{nd} 
order yields the time-step operator we use.	 For problems involving very high frequencies, higher order Magnus expansions are advantageous,
like the 4\textsuperscript{th} order expansion $\exp(-i\bar{H}(t_1)\Delta t -i\frac{1}{2}\Delta t(\hat{H}(t_1)+\hat{H}(t_2)) - 
\frac{\sqrt{3}}{12}\Delta t^2[\hat{H}(t_2),\hat{H}(t_1)] )$, where $t_{1,2}=t+(\frac{1}{2} \mp \frac{\sqrt{3}}{6}) \Delta t$. 
However, this is beyond our scope.}.                          
This way it becomes extremely easy to handle the time-dependence of the Hamiltonian $\hat H(t)$,
since for the n\textsuperscript{th} time-step we simply use the midpoint Hamiltonian $\bar{H}(t_n) = \hat H(\frac{1}{2}(t_{n-1}+t_n))$,
as illustrated in \cref{fig:strategy}~A. The price we pay is that we have to apply the time-step operator a lot of times. However, since $\Delta t$ is 
small, we can use simple and effective approximations for the time-step operator (as explained in \cref{sec:TSO}), leaving the numerical 
cost reasonable.

The above strategy is the obvious choice numerically, and works very well, for the case where the Hamiltonian does not depend 
on the wave function. However, in our case, where the Hamiltonian depends on the wave function $\ket|\Psi(t)>$ 
(and target $S(t)$), i.e., $H[\Psi,S]$, we are unable to calculate the midpoint Hamiltonian,  since we do not
 know the midpoint wave function a priori.

The simplest strategy in this case is then to just use the on-point Hamiltonian $\hat H(t_n)$ instead of the midpoint Hamiltonian $\bar H(t_{n+1})$ 
(as indicated in \cref{fig:strategy}~B above the arrows). However, this is clearly not very precise, or even physical, as it breaks the 
time-reversal symmetry. Thus it requires relatively small time-steps to work, leaving it numerically inefficient.
However, having calculated $\ket|\Psi(t_{n+1})>$ approximatively, with the help of the on-point Hamiltonian $\hat H(t_n)$
we can obtain an approximate midpoint Hamiltonian,	 which can then be used to improve $\ket|\Psi(t_{n+1})>$.												
This is an effective way to get most of the ''midpoint effect''. Note that it is also possible to only use midpoint
Hamiltonians in this situation, as illustrated in \cref{fig:strategy}~C. However, this strategy requires an equidistant time-grid, 
preventing (much more efficient) adaptive time-steps. To initialize this strategy, we compute $\ket|\Psi(t_0)>$ and 
$\ket|\Psi(t_1)>$ by propagating $\ket|\Psi_0>$ using the Hamiltonian $\hat{H}(t_1)$, instead of the midpoint Hamiltonians 
(just as in \cref{fig:strategy}~B)\footnote{While this initialization is not as good as the actual propagation, it is only for one 
time-step out of thousands, so it is a small issue.}.
We can then calculate the ''midpoint'' Hamiltonian for the steps of length $2\Delta t$, e.g. $\hat H(t_1)$.
So this is really a clever way to solve the issue of calculating the Hamiltonian.						
Note however, that only even, respectively odd times $t_n$ are connected to each other,
leading to somewhat different errors of even and odd $\ket|\Psi(t_n)>$, and hence slightly oscillatory 
expectation values.  Yet another alternative for this type of wave-function-dependent Hamiltonians is to
extrapolate the Hamiltonian to the midpoint (in \cref{fig:strategy}~A one would thus use the extrapolated 
midpoint Hamiltonian) \footnote{Strategies B, C and the extrapolation method apply equally well to Kohn-Sham theory,			
where the Hamiltonian depends on the wave function through the density (also on previous times, although in practise this 
dependence is usually ignored).}.

Strategies B, C and extrapolation usually work well for LCT, however, in the case of density control this is no longer true, as it is much harder to stabilise.
In this case, we therefore need an entirely different time-stepping strategy that focus on the fact that we have a control 
target that we want to achieve.

We start by making an initial guess for the midpoint Hamiltonian $\bar{H}_0(t_{n+1})$, e.g. say by (linear) extrapolation.
Provided we can come up with an update formula,												
\begin{equation} \label{updateFormulaH}
\bar H_{k+1}(t_{n+1}) = \bar H_k(t_{n+1}) + f(S_k(\br t_{n+1}) - S(\br t_{n+1})) ,
\end{equation}
that corrects the midpoint Hamiltonian based on how far the actual path $S_k(\br t_{n+1})$ is from the prescribed path $S(\br t_{n+1})$,
i.e. based on the residual $S_k(\br t_{n+1}) - S(\br t_{n+1})$, we can repeat this until both paths practically coincide,
and move to the next time-step, as illustrated in \cref{fig:strategy}~D.
Note that, since in the Hamiltonian only the external field may change, we may equivalently rewrite \cref{updateFormulaH} as
\begin{equation} \label{updateFormula}		
\bar v_{k+1}(\br t_{n+1}) = \bar v_k(\br t_{n+1}) + f(v_k(\br t_{n+1}) - v(\br t_{n+1})). 
\end{equation}
The huge advantage of this strategy is that it always tries to compensate any error in the time-propagation 
by changing the midpoint Hamiltonians to ensure that the path stays correct.
Thus, even if $S(\br t)$ changes by orders of magnitude, as in our case, it will likely still get it right
(while the potential we find is not quite equal the exact). Using any of the above strategies any error in $S(\br t)$ 
will never be corrected, except by sheer luck, and what may be a small error to start with, may become a major error
if $S(\br t)$ changes by orders of magnitude, causing a break down.
In short, strategy D is generally slightly more expensive, but much more stable. We shall also see in \cref{sec:DLCT}, 
that it easily allows nice further stabilising techniques, like ensuring that the continuity equation holds.
In our case, the update formula \labelcref{updateFormula} is also much simpler than \cref{nvq}, allowing 
for a simpler implementation. 

In the appendices \ref{sec:TSO} and \ref{sec:SR}  we discuss further numerical aspects of time-propagation.

%--------------------------------------------------------------------------------------------------------------------------------------------------%
\section{Density Local-Control Theory} \label{sec:DLCT}						
%--------------------------------------------------------------------------------------------------------------------------------------------------%

To make use of our new time-stepping strategy, we need to derive an update formula \labelcref{updateFormula}.
To do this we first note that \cref{nvq} can be solved iteratively.
To do so, we define an iterative sequence $v_k$ of potentials by the iterative solution of
\begin{equation} \label{fp}
-\nabla \cdot \left( n(\br t)\nabla v_{k+1}(\br t) \right) = q([v_k] , \br t) - \partial_t^2 n(\br t) .
\end{equation}
In previous works \cite{GFPP,GFPP1D} we showed, for general initial states and interactions,
that under mild restrictions on the density the sequence $v_k$ converges in Banach-norm sense 
to a potential $v$ that produces the prescribed density $n$ and is the fixed-point of the equation.
To obtain an equation that only depends on densities, we use \cref{nvq} for a system with 
potential $v_k$ to eliminate the quantity $q$ from \cref{fp},
\begin{align} \label{fpn}
-\nabla \cdot \left( n(\br t)\nabla v_{k+1}(\br t) \right) &= \partial_t^2 \left[ n([v_k],\br t) - n(\br t) \right] \nonumber \\
&-\nabla \cdot \left( n([v_k],\br t)\nabla v_k(\br t) \right) .
\end{align}
As $v_k$ approaches $v$, $n[v_k]$ approaches $n$, so we can replace the last $n[v_k]$	by $n$ in the near convergence region,
\begin{equation} \label{fpnsimple}
-\nabla \cdot \left( n(\br t)\nabla \left[ v_{k+1}(\br t) - v_k(\br t) \right] \right) =
\partial_t^2 \left[ n([v_k],\br t) - n(\br t) \right] .
\end{equation}
Since we can invert the operator $-\nabla \cdot n(\br t)\nabla$ this 
shows us how we can update $v_k$  given the residual $n[v_k]-n$, and (once time-discretized) is  
our desired update formula.

To also make use of the current (which will help us in stabilising the numerics), we note that using the continuity 
equation this may also be written as						
\begin{align} \label{fpjsimple}
\nabla& \cdot \left( n(\br t)\nabla \left[ v_{k+1}(\br t) - v_k(\br t) \right] \right) \nonumber
\\
&= \partial_t \left[ \nabla \cdot \mathbf{j}([v_k],\br t) + \partial_t n(\br t) \right] .
\end{align}

Note that \cref{fpnsimple} only is converged if $v_k$ really yields $n$,
while \cref{fpjsimple} first is converged when $v_k$ really yields a $\mathbf{j}[v_k]$, that together with $n$ satisfy the continuity equation.
By combining these, we require both $n$ and $\mathbf{j}$ to be as correct as possible, even when numerical errors try to build up. Consequently we 
find
\begin{align} \label{fpnjsimple}
-\nabla \cdot &\left( n(\br t) \nabla \left[ v_{k+1}(\br t) - v_k(\br t) \right] \right) = \nonumber 
\\
&(1-\mu) \partial_t^2 \left[ n([v_k],\br t) - n(\br t) \right] - 
\\
&\mu \partial_t \left[ \nabla \cdot \mathbf{j}([v_k],\br t) + \partial_t n(\br t) \right] , \nonumber
\end{align}
where $\mu$ is a parameter at our disposal.

Finally, we implement the update formula \cref{fpnjsimple} stepwise in time\footnote{Since we iterate the update formula for every time step, the 
converged potential of the previous time step is a very good initial guess for the potential at the present time step. This justifies that we employ the simplified update formula
where we used $n[v_k] \simeq n$. Even in the initial time step (where we have no previous potential to take as initial guess) we never experienced 
any problems with this.}.  By denoting $\Delta t = t_{n+1}-t_n$, using only times $t_m$ with $m \le n+1$ for the derivatives and employing the fact that we have 
already converged up to time $t_n$ we find from \cref{fpnjsimple} by a discretization with respect to time 
\begin{align} \label{fpnjsimplenum}
&- \nabla \cdot \left( \bar{n}(\br t_{n+1}) \nabla \left[ \bar{v}_{k+1}(\br t_{n+1}) - \bar{v}_k(\br t_{n+1}) \right] \right) \Delta t^2  \nonumber \\
&= A \left[ n([v_k],\br t_{n+1}) - n(\br t_{n+1}) \right] \nonumber \\
&- B \Delta t \left[ \nabla \cdot \mathbf{j}([v_k],\br t_{n+1}) + \partial_t n(\br t_{n+1}) \right] .
\end{align}
The constants $A$ and $B$ depend on the discretization scheme and the $\mu$ of \cref{fpnjsimplenum},
which effectively leaves the choice of their values at our disposal. In practise we generally employ values between $0.5$
and $1$.

All that is left to do is to find a way to solve the above Sturm-Liouville problem, i.e, invert the self-adjoint operator 
$-\nabla \cdot n(\br t) \nabla$. To do so we need to choose boundary conditions
on $v$ which are in accordance to the quantum system we want to look at (see \cref{sec:BC}).  After discretisation 
with respect to space (see \cref{sec:SR}) we can then for example use relaxation methods (possible accelerated by multi-grid methods for very high efficiency) 
to solve the resulting inversion problem. This allows us to also consider two- and three-dimensional quantum systems, 
while in our previous work \cite{Nielsen2012} we were restricted to one-dimensional multi-particle problems, where a direct integration 
of the Sturm-Liouville operator is possible\footnote{There the choice of appropriate boundary conditions 
give rise to non-trivial integration constants \cite{Nielsen2012}.}.

The resulting algorithm usually converges within 5 to 10 iterations for each time-step, making it 
essentially as fast as the time-propagation scheme.

%--------------------------------------------------------------------------------------------------------------------------------------------------%
\section{TDDFT Version} \label{sec:TDDFT}				
%--------------------------------------------------------------------------------------------------------------------------------------------------%
																					
While the stepwise approach reduces the cost of our method to a few times that of a normal 
time-propagation, it is still way to expensive to apply to anything larger than a few electrons.
However, our method apply equally well to non-interacting and interacting systems, and in this 
section we will show that by applying our method to an auxiliary non-interacting system,
we can actually obtain the external potential of the interacting system by just subtracting a 
TDDFT approximation.

For a non-interacting system, the wave function $\ket|\Phi(t)>$ can be represented by orbitals $\phi_i(\br t)$,		
provided the initial state $\ket|\Phi_0>$ is given in terms of orbitals. To do the time-propagation in this case,
we thus only need to propagate each individual orbital $\phi_i(\br t)$ using
\begin{equation*}																		
\hat H(t) = - \tfrac{1}{2} \nabla^2 + v_s(\br t),
\end{equation*}
where we employed the usual convention to denote an external potential that acts on a 
non-interacting system by $v_s$. The density can then also be expressed in terms of the orbitals by
\begin{equation*}	
n(\br t) = \sumN |\phi_i(\br t)|^2 .
\end{equation*}
Thus time-propagation, and therefore also our method, are much less numerically demanding 
for non-interacting systems. This fact is used in the Kohn-Sham approach to TDDFT, where one 
uses an auxiliary non-interacting system to predict the density of an interacting reference system.
To force the non-interacting Kohn-Sham system to have the same density as the interacting one,
an auxiliary field the so-called Hartee-exchange-correlation (Hxc) potential 
\begin{equation}\label{HxcPotential}	
v_\mathrm{Hxc}[\Psi_0,\Phi_0,n] = v_s[\Phi_0,n]-v[\Psi_0,n],
\end{equation}
is introduced (see \cite{CarstenBook, TDDFT} for details). Its individual terms $v_s[\Phi_0,n]$ and $v[\Psi_0,n]$ are the control fields 
that force the different quantum systems, starting from $\ket|\Phi_0>$ and $\ket|\Psi_0>$ respectively, to produce
the same density $n(\br t)$ via propagation.  The Hxc functional is then usually rewritten as 
\begin{equation*}
v_\mathrm{Hxc}[\Psi_0,\Phi_0,n] = v_\mathrm{H}[n] + v_\mathrm{xc}[\Psi_0,\Phi_0,n]
\end{equation*}
where the Hartree potential (i.e., the classical interaction contribution)
\begin{equation*}
v_\mathrm{H}([n], \br t) = \int d \br' n(\br' t) w(|\br - \br'|)
\end{equation*}
is made explicit. The resulting unknown part is called as the exchange-correlation (xc) potential
$v_\mathrm{xc}[\Psi_0,\Phi_0,n]$ and is an intensively studied quantity in TDDFT, for which 
decent approximations exist. Thus, to calculate the control field that generates for a given
initial state $\ket|\Psi_0>$ (in the interacting system) a prescribed density $n$, i.e., $v[\Psi_0,n]$, 
we simply calculate $v_s[\Phi_0,n]$ for some initial state $\Phi_0$ (having the same initial density and time-derivative 
of the density as the interacting problem, usually found by ground-state DFT),
and then we subtract a TDDFT approximation $v_\mathrm{xc}^{\mathrm{app}}$ to obtain $v[\Psi_0,n]$, i.e.,
\begin{equation} \label{ApproximateField}	
v[\Psi_0,n] \simeq v_s[\Phi_0,n] -  v_\mathrm{H}[n] - v_{\mathrm{xc}}^{\mathrm{app}}[\Psi_0,\Phi_0,n].
\end{equation}
How this works in practise for a simple approximation (which ignores the dependence on 
$\ket|\Phi_0>$ and $\ket|\Psi_0>$ as is usually the case for most approximate xc potentials ) will be 
shown in \cref{sec:Examples}.

%--------------------------------------------------------------------------------------------------------------------------------------------------%
\section{Optimal Control Theory} \label{sec:OCT}				
%--------------------------------------------------------------------------------------------------------------------------------------------------%

Our density-control method also allows us to do OCT, i.e., to find the path that optimizes the time-integral 
of the expectation value of a possibly time-dependent observable $\hat O(t)$\footnote{In principle we allow 
for a sum of time-dependent observables which can depend themselves on the wave function and can give 
different times a different importance by multiplying with a weighting function.} 
over a fixed time interval $[t_0,T]$, i.e., we search for the $\ket|\Psi(t)>$ that optimises
\begin{equation*}	
J[\Psi]= \int_{t_0}^{T } d t \, \bra< \Psi(t) |\hat O(t) \ket|\Psi(t)>
\end{equation*}
for a fixed initial state. To find this optimal path we have to search over all wave functions that satisfy 
the TDSE (which in standard OCT is done using a Lagrange multiplier to ensure that $\ket|\Psi(t)>$ 
satisfies the TDSE). This is extremely involved since even algorithms optimized for this will have to update 
$\ket|\Psi(t)>$ thousands of times before the search approaches the optimal wave function \cite{OCT1},
and each of these updates involves a time-propagation of the full many-body state. The search would be 
much simpler (and thus faster) if we could somehow restrict it to only  a small class of relevant wave functions,
which is exactly what our algorithm offers. Basically, it is hard to restrict the time-dependent wave functions 
explicitly, since they must satisfy the TDSE and we have little intuition about these states. It is much easier in
the case of the densities. Since most time-dependent densities are valid (if they have the same density and 
time-derivative of the density as the initial state and are smooth enough in time and space 
\cite{vanLeeuwen, GFPP, GFPP1D}) we can apply our physical intuition to restrict to sensible density profiles.

%--------------------------------------------------------------------------------------------------------------------------------------------------%

To actually do this density-based OCT, we first employ the fact that we can label all wave functions by
their respective density (which is a consequence of the Runge-Gross theorem \cite{RungeGross}) $\ket|\Psi([\Psi_0,n],t)>$.
Thus, instead of optimizing with respect to the wave function we can equivalently optimize with respect to the density, i.e.,
\begin{equation*}	
J[\Psi_0, n]= \int_{t_0}^{T } d t  \, \bra< \Psi([\Psi_0,n],t) |\hat O (t) \ket|\Psi([\Psi_0,n],t)>.
\end{equation*}
Of course, the full optimization over all possible densities is numerically as expensive as the optimization with 
respect to the wave functions. However, if we restrict the set of allowed densities we can greatly reduce the
numerical costs of such an optimization. For instance, if we want to minimize the required field energy 
$J[n]=\int_{t_0}^{T } d t \int d \br (\nabla v([\Psi_0,n],\br t))^2$ for a rigid translation of the initial density
to some specific point at $t=T$, we can use linear combinations of densities that all will arrive at the specific 
point at $t=T$. How this can be done in practice is shown in \cref{sec:Examples}. 

Of course, we can combine this approach to OCT with the ideas introduced in the previous section. Instead of making an
optimization in the fully interacting quantum system we can consider the non-interacting Kohn-Sham system.
However, this requires that we can express $J$ of the interacting system in terms of the density or the orbitals
of the Kohn-Sham system, which can rarely be done exactly (see for a short list of such cases below). Thus we face 
the same challenges as standard OCT when employed together with TDDFT, i.e., we usually only have crude approximations
to the observables of the interacting system.

This brings us to another possibility of combining local density control with OCT. In the special case where we can 
express, or adequately approximate, the functional in terms of the density only,		
we can split the OCT problem up into finding an optimal density and then use tracking,
thus completely avoiding the extremely expensive optimization in terms of the wave functions.
This includes the important class of observables based on the spatial operator $\hat{\br}$,						
since the expectation value of such an operator $\hat f(\hat {\br})$ is given by $f(t) = \int f(\br) n(\br t) d\br$.
For instance, the dipole acceleration, which is proportional to the radiation of the quantum system, is	given by					
$\ddot{\bf \mu}(t) = \int \br \ddot{n}(\br t)d\br $.
It also includes the particle-number operator restricted to a specific region, e.g., around the initial 
location of the system ($f(\br) = 1$ near 
the initial location and zero elsewhere), which approximates the number of bound and ionized 
electrons (since both have to add up to $N$) \cite{CarstenBook}. Further, also the divergence of the current $\mathbf{j}$ and the
local force $\mathbf{Q}$ are given in terms of the density through the continuity equation and local-force equation. 
Unfortunately, there are not many other cases\footnote{Although the Runge-Gross theorem \cite{RungeGross} 
proves that every observable is a unique functional of the density and initial state,
the explicit form of these functionals is only known in very rare cases.}. An analytical example
of such an explicitly density based optimization can also be found in \cref{sec:Examples}.

We finally emphasize that one can employ these optimal control ideas also with other local-control schemes, such 
as dipole tracking, provided that we only have $\varepsilon(t)$ to vary. In other cases, such as population tracking, we cannot
combine our LCT and OCT straightforwardly since we do not know the set of valid time-dependent populations.

%--------------------------------------------------------------------------------------------------------------------------------------------------%
\section{Examples} \label{sec:Examples}
%--------------------------------------------------------------------------------------------------------------------------------------------------%

To illustrate the capabilities of our numerical procedure, we focus on a simple model system (in 1D and 2D), 
that is easy to introduce. This lets us keep the focus on what is actually possible, instead of details of specific 
systems. While 3D systems also are feasible with our method, we refrained from considering 
those in this work since they are harder to illustrate. In particular, the following examples show that our algorithm
converges for large and rapid density changes of orders of magnitude irrespective of the initial state, the interactions
and the dimensionality of the quantum system.

%--------------------------------------------------------------------------------------------------------------------------------------------------%
\subsection{1D Model System}
%--------------------------------------------------------------------------------------------------------------------------------------------------%

In all these cases, we consider $N$ electrons on a quantum ring of length $L=10$ over a time interval of length $T=20$.	
We start by calculating an initial state $|\Psi_0 \rangle$, which in all cases is a ground state or excited state	     
of a (properly periodic) Hamiltonian with external potential $v_0$ and interaction $w$ given by						
\begin{align*}
v_0(x) &= - \cos \left( \frac{2\pi x}{L} \right) , \\
w(x_1,x_2) &= \lambda \cos \left( \frac{2\pi(x_1-x_2)}{L} \right) ,
\end{align*}
where $\lambda$ is the interaction strength.
The initial density is denoted by $n_0 (x)$.												
We then construct the (spatially periodic) time-dependent densities $n_1(xt)$ and $n_2(xt)$ by:						
\begin{align*}
n_1 (x t ) &= n_0 (x-r(t)), \\
n_2 (x t)  &= \tfrac{1}{2} [ n_0(x-r(t)) + n_0(x+r(t))], \\
r\left( t \right) &= \frac{L}{2}\left[ {1 - \cos \left( \frac{\pi t}{T} \right)} \right].
\end{align*}
The density $n_1$ describes a situation where the initial density $n_0$ is rigidly translated around the ring exactly once,
whereas the density $n_2$ describes a situation where the initial density $n_0$ is split in equal halves $\frac{1}{2} n_0$
that are rigidly translated in opposite directions to rejoin at times $\frac{1}{2} T$ and $T$. These examples are very 
challenging, since the external potential has to move the densities by orders of 
magnitude (for a fixed position $x$), and, in the case of a splitting, ensure the wave function splits and recombines correctly. We have then used our algorithm to 
calculate the potentials that	produce these prescribed densities $n_1$ and $n_2$ via time-propagation of the initial state 
$| \Psi_0 \rangle$ by the TDSE. This was done for the interaction strengths $\lambda=1$ and $\lambda=0$, 
corresponding to interacting and non-interacting situations below, respectively.

\subsubsection{Interacting} \label{sec:Example1DModelInteracting}
\begin{figure} [H]
\includegraphics[width=8.6cm]{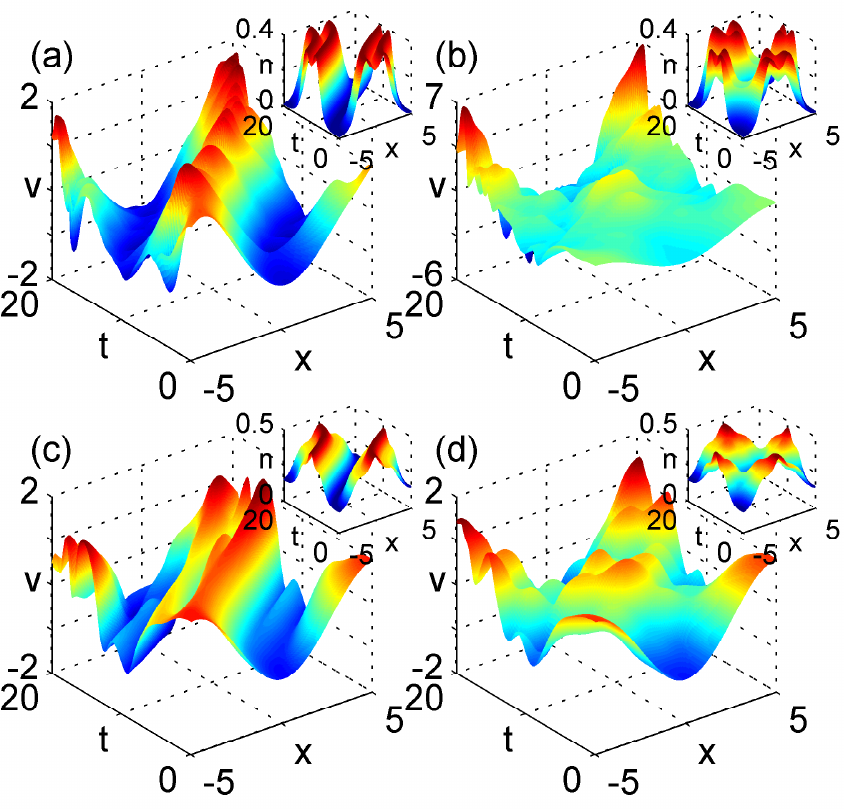}
\caption{(color online) The potentials (including the static $v_0$)  that (a) translate and (b) split the 2-electron density and that (c) translate
and (d) split the 3-electron density.}
\label{fig:I1Potentials}
\end{figure}

Here we consider 2 cases. In the first case we take the initial state $|\Psi_0 \rangle$ to be the 2 electron singlet ground state\footnote{We
have considered this example already in \cite{Nielsen2012}. However, since we will use it later to benchmark TDDFT approximations we present it also here.} 
(which has a symmetric spatial wave function), while in the second case we take the initial state to be the 3 electron state
 with all spin up (or down) (which has an anti-symmetric spatial wave function). The corresponding potentials (including the static potential $v_0$) 
 and densities (insets) for both cases are shown in \cref{fig:I1Potentials}.

\subsubsection{Non-Interacting}  \label{sec:Example1DModelNonInteracting}
\begin{figure} [H]
\includegraphics[width=8.6cm]{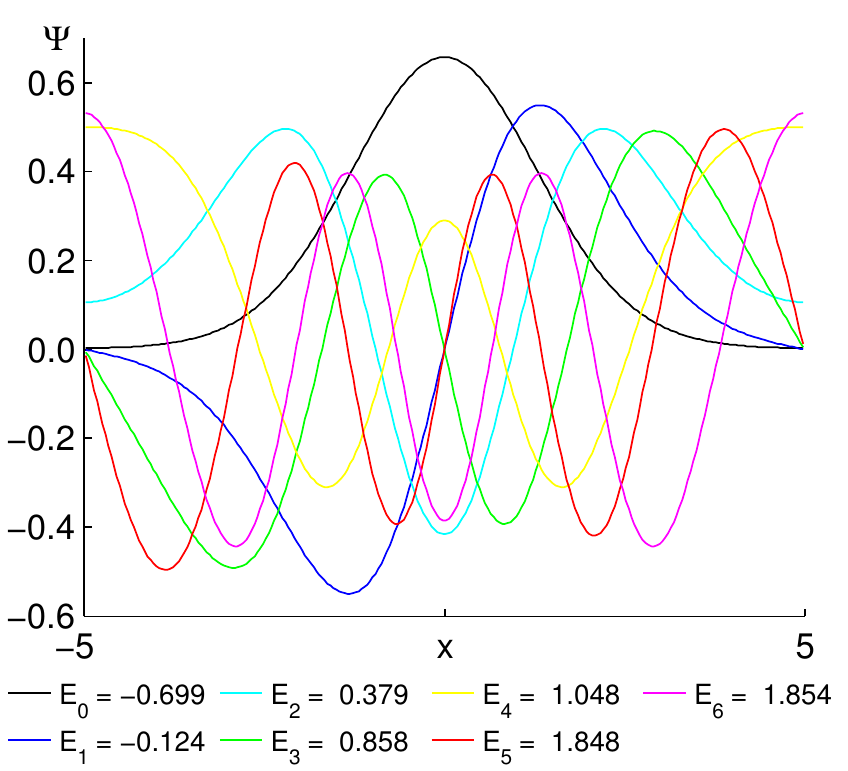}
\caption{(color online) The first seven one-electron eigenstates of $v_0$.}
\label{fig:orbitals}
\end{figure}

\begin{figure} [H]
\includegraphics[width=8.6cm]{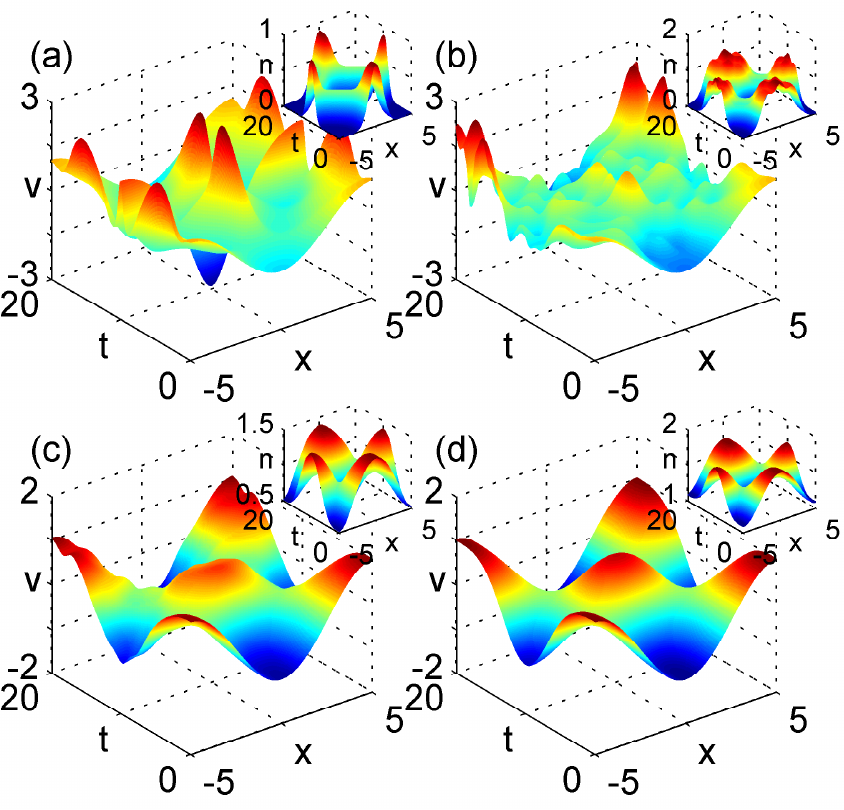}
\caption{(color online) The potentials (including the static $v_0$)  that split (a) the 2- (b) 6- (c) 10- and (d) 14-electron density.}
\label{fig:NIPotentials}
\end{figure}

Here we consider 4 different cases of splitting, where we take the initial states to be the 2, 6, 10 and 14 electron ground states\footnote{In the cases
of 4 and 8 particles the algorithm becomes unstable. This is because the excited state orbitals dominate the density near their
nodes and extremely strong potentials are needed to control the density when it goes to zero. }.
Since the electrons do not interact, we can construct these ground states by computing the one-electron eigenstates of \cref{fig:orbitals},		
and use them as initial orbitals by the Aufbau principle. We then do the splitting, to get the potentials of \cref{fig:NIPotentials}. 
Note how we can treat much larger non-interacting systems, than interacting,
due to the linear scaling of non-interacting problems.
																	      
\subsubsection{Interacting by Non-Interacting}						

To see how we can take advantage of the fact that we numerically can treat much larger non-interacting systems than interacting, we employ the idea presented in \cref{sec:TDDFT}. As an example,
we  approximate the splitting of the interacting 2-electron problem of \cref{sec:Example1DModelInteracting} by 
working with an auxiliary non-interacting system. While this procedure is of course intended to approximate the control field
of large interacting systems, by considering a simple case instead, we can easily test how well this idea works in practise.

To do this, we need to employ an approximation to the unknown functional $v_\mathrm{\mathrm{xc}}[\Psi_0,\Phi_0,n]$ as 
discussed in \cref{sec:TDDFT}.  While in principle these functionals depend on the initial states
of the interacting and the non-interacting system, in practise one tries to avoid these formal dependences. This can be done
if one starts the propagation from interacting and non-interacting ground states that have the same density \cite{CarstenBook, TDDFT}. In this case
$v_\mathrm{\mathrm{xc}}[n]$ only depends on the density and we can use ground-state DFT \cite{DFT} to determine an
approximation to the ground-state density. Thus, strictly speaking, we have two different approximations in this DFT scheme.
One that approximates the initial density (or equivalently the ground-state Kohn-Sham potential) and one that approximates
$v_\mathrm{\mathrm{xc}}[n]$. As almost always done in TDDFT we will employ for both problems the same approximation.
In the special case of two electrons in a singlet state, the exact exchange approximation to $v_\mathrm{\mathrm{xc}}[n]$ is particularly 
simply, and hence an obvious choice. Indeed, in this case, this approximation is just half the Hartree potential, i.e.,
 $v_\mathrm{\mathrm{xc}}[n] \simeq -\frac{1}{2} v_\mathrm{H}[n]$ \cite{CarstenBook}.

\begin{figure} [H]
\includegraphics[width=8.6cm]{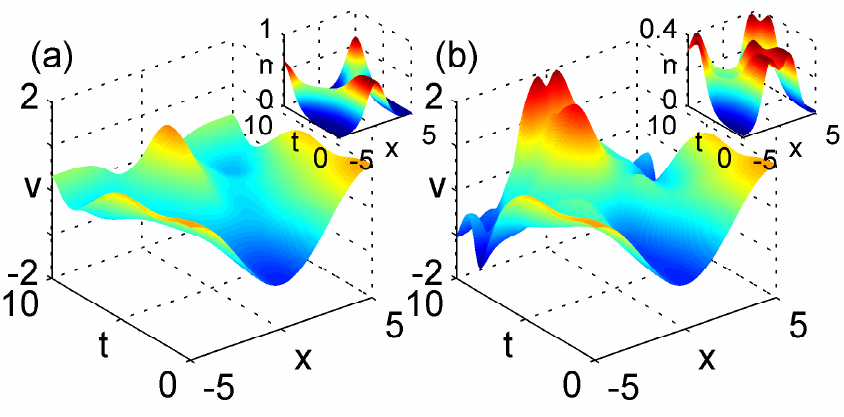}
\caption{(color online) (a) Approximate potential and density for the exact-exchange approximation and (b) exact potential and density in 
the case of a 2-particle spin-singlet problem. Note that we only plotted $t \in [0,10]$.} 
\label{fig:PotentialI1ByNI}																			  
\end{figure}

From this we can determine the approximate Kohn-Sham ground state $\ket|\Phi_0>$, which in turn approximates 
the ground state density of the interacting 2-particle problem of  \cref{sec:Example1DModelInteracting}.
Given these functions we simply prescribe $n_2(xt)$ based on the approximate density
and compute the external potential $v_s(xt)$ that yields this density from the initial state $\ket|\Phi_0>$,
just like for the non-interacting systems in \cref{sec:Example1DModelNonInteracting}.
Subtracting the exact exchange approximation $v_\mathrm{Hxc}[n_2] \simeq \frac{1}{2} v_\mathrm{H}[n_2]$ 
for the density $n_2(xt)$, we then obtain our approximation to $v[\Psi_0,n_2]$, 
which we compare to the exact in \cref{fig:PotentialI1ByNI}. 

We see that the approximation works well for $t \in [0,5]$, but then starts to deviate as the dynamics gets stronger.
After $t=10$ the approximate potential is far off, since it is essentially symmetric about $t=10$ and hence almost stays 
between -1 and 1, while the exact is rather between -6 and 7 as seen in \cref{fig:I1Potentials}.
So the combination with TDDFT approximations allows us get a some reasonable information on the control field of an interacting systems,
at least for some time. Finally we then consider how well the approximate control field works for the 
interacting system, i.e., we apply the potential of \cref{fig:I1Potentials}~(a) to the interacting 2-particle system of 
\cref{sec:Example1DModelInteracting}. We find in \cref{fig:PotentialI1ByNI}~(b) that although the density 
disperses (and thus does not keep its shape) a splitting of the initial density is clearly visible, although it does not properly 
rejoin at $T$. Therefore, the TDDFT approximation allows to find a reasonable initial guess for a control field of
an interacting system.

\begin{figure} [H]
\includegraphics[width=8.6cm]{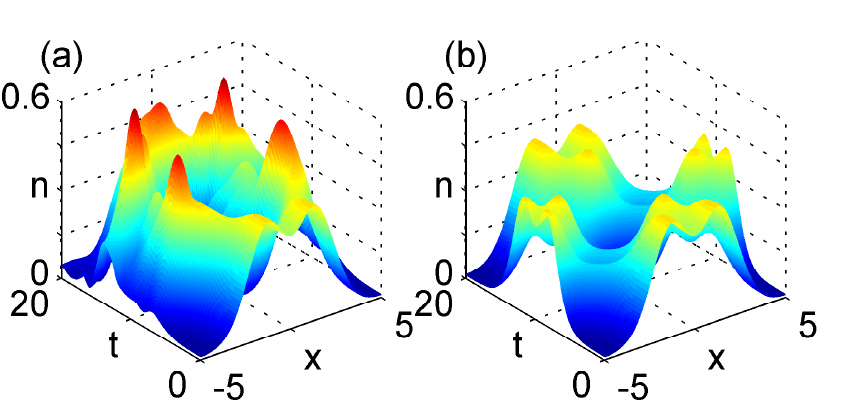}
\caption{(color online) (a) Density based on the exact-exchange approximated control field (b) the density from the exact control field.}
\label{fig:DensitiesI1ByNI}
\end{figure}

\subsubsection{Benchmarking Exchange-Correlation Approximations}						

To apply the approach of \cref{sec:TDDFT} to larger systems, or to use TDDFT in general,					
users need some knowledge of how well different approximations work in various cases.
Therefore, mainly for developers of $v_\mathrm{Hxc}$ approximations,
we will here give an example of how one can construct the exact $v_\mathrm{Hxc}$			
for very small systems, which can then be used to benchmark the various approximations,								
and provide novel insight into the properties of the exact Hxc potential.

To do this, we first compute the exact Kohn-Sham initial state $\ket|\Phi_0>$ corresponding to the ground state
density of $\ket|\Psi_0>$, which is itself the ground state of the non-interacting Kohn-Sham Hamiltonian 
for the Kohn-Sham ground-state potential $v_s[n_0]$.  A very simple but not most efficient and stable way 
to do this is to start with an initial guess $v^0$ for the potential and compute the corresponding non-interacting 
ground state density $n^0$. Then we update the potential by 
$v^{i+1} = \frac{n^i}{n_0} v^i$, with $n_0$ the density we want, until convergence $v^i \to v_s[n_0]$ and $\Phi^i \to \Phi_0$.

\begin{figure} [H]
\includegraphics[width=8.6cm]{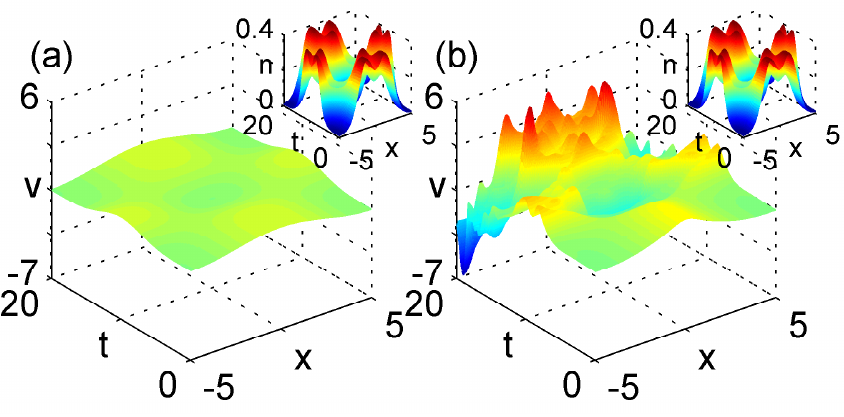}
\caption{(color online) (a) Approximate (exact-exchange) (b) and exact Hxc potential for the splitting
of the interacting 2-particle system.}
\label{fig:vHxc}
\end{figure}

In a next step we then determine the potential  $v_s[\Phi_0, n]$ that generates the density of the interacting problem (in this case
the density $n_2$ of the 2-particle problem of \cref{sec:Example1DModelInteracting}) in the non-interacting system
and subtract the potential $v[\Psi_0, n]$ of the interacting system, i.e., \cref{HxcPotential} . This determines the exact 
Hxc potential for this case. The approximate Hxc potential is then determined by plugging the exact density into the 
approximate functional, in our case we consider $v_{\mathrm{Hx}}[n_2] = \frac{1}{2} v_\mathrm{H}[n_2]$. How they 
compare is displayed in \cref{fig:vHxc}. Similar to our previous examples, we see that the approximation is reasonable for 
roughly half of the time interval.  However, while the exact-exchange approximation is ``adiabtic'' (it only depends on the
density at the same time), and thus by construction symmetric around $t=10$, the exact Hxc potential depends on all 
previous times. Therefore this strong deviation is a clear indication of the non-adiabaticity of the Hxc potential in this
problem.  

%--------------------------------------------------------------------------------------------------------------------------------------------------%
\subsection{2D Model System}
%--------------------------------------------------------------------------------------------------------------------------------------------------%

To illustrate that the method also works the same in two dimensions,  we consider an analogous 2D model system, 
for which \footnote{This interaction is a bit unusual since it depends on the direction,
and thus is different for $x_1-x_2=\sqrt{2}$, $y_1-y_2=0$ and $x_1-x_2=1$, $y_1-y_2=1$. 
However, the algorithm can also handle such a case.}									
\begin{align*}
v_0(x,y) &=  - \cos \left( \frac{2\pi x}{L} \right) - \cos \left( \frac{2\pi y}{L} \right) , \\
w(x_1,y_1,x_2,y_2) &= \lambda \cos \left( \frac{2\pi(x_1-x_2)}{L} \right) 
\\
&+ \lambda \cos \left( \frac{2\pi(y_1-y_2)}{L} \right) ,
\end{align*}
and where we still take $L=10$ (both sides) and $T=20$, and still use periodic BCs.
We then compute an initial state as before, and construct a time-dependent density
\begin{align*}																				
n(xyt) = \tfrac{1}{4} &[ n_0(x-r(t),y) + n_0(x+r(t),y)\\
				&+ n_0(x,y-r(t)) + n_0(x,y+r(t)) ],
\end{align*}
which describes a splitting in 4, instead of just 2 as before\footnote{If we do a translation or splitting into 2 parts we can get the 2D potentials
also from a 1D calculation, since one direction is then constant. This can be used to verify that the algorithm works equally well in one or two
dimensions}. We employ the same $r(t)$ as in the 1D case and again use our algorithm to compute $v$.

\subsubsection{One Particle}

\begin{figure} [H]
\includegraphics[width=8.6cm]{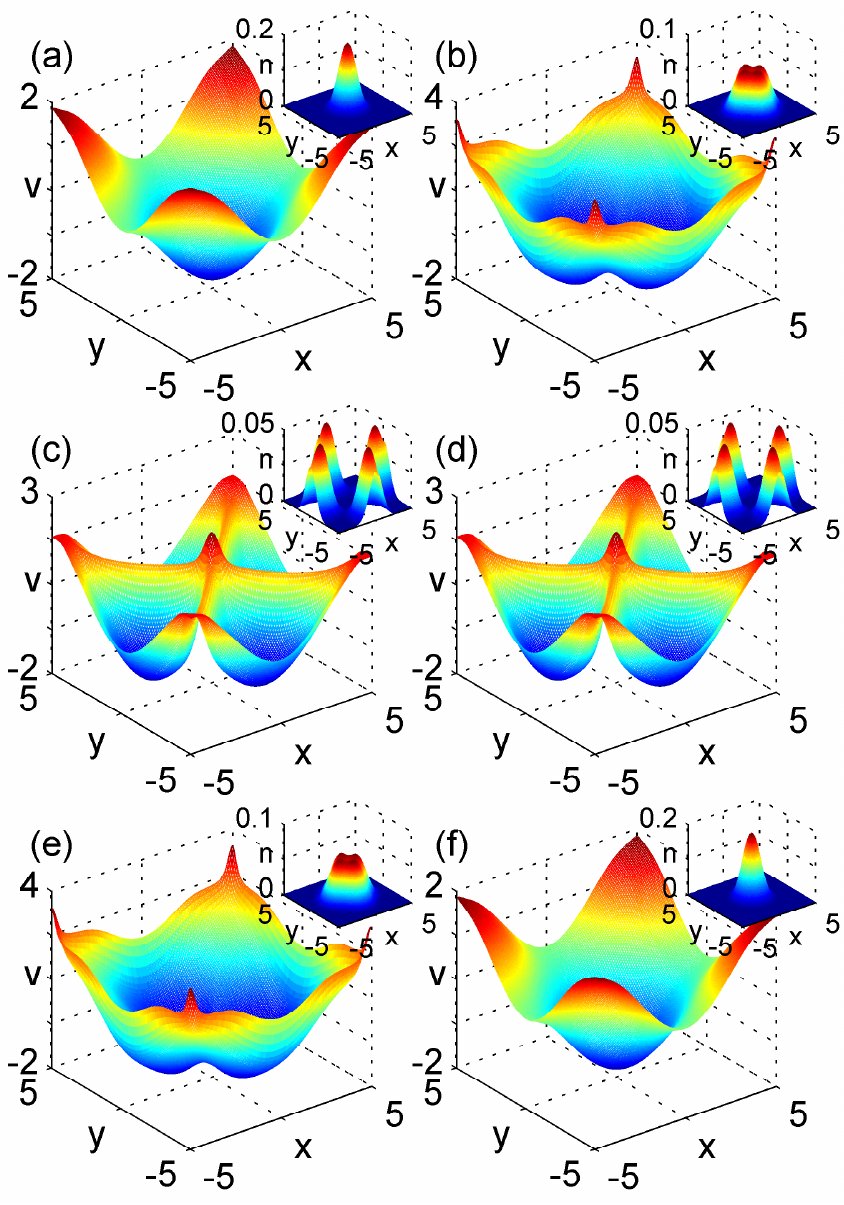}
\caption{(color online) Snapshots of the potential and splitted density (insets) for a single 2D particle at different times. (a) t=1,67, (b) t=5,
(c) t=8,33, (d) t=11,67, (e) t=15 and (f) t=18,33. }
\label{fig:oneSplitIn4}															
\end{figure}

In our first case, we simply consider a single electron, and take the initial state to be the ground state of the above potential $v_0$	.
Snapshots of the resulting two-dimensional potential and the respective (splitted) density are given in \cref{fig:oneSplitIn4}. While a representative
presentation of the two-dimensional potential would need much more snapshots, the pictures illustrate that the potential changes
a lot throughout time.

\subsubsection{Many Particles}

\begin{figure} [H]
\includegraphics[width=8.6cm]{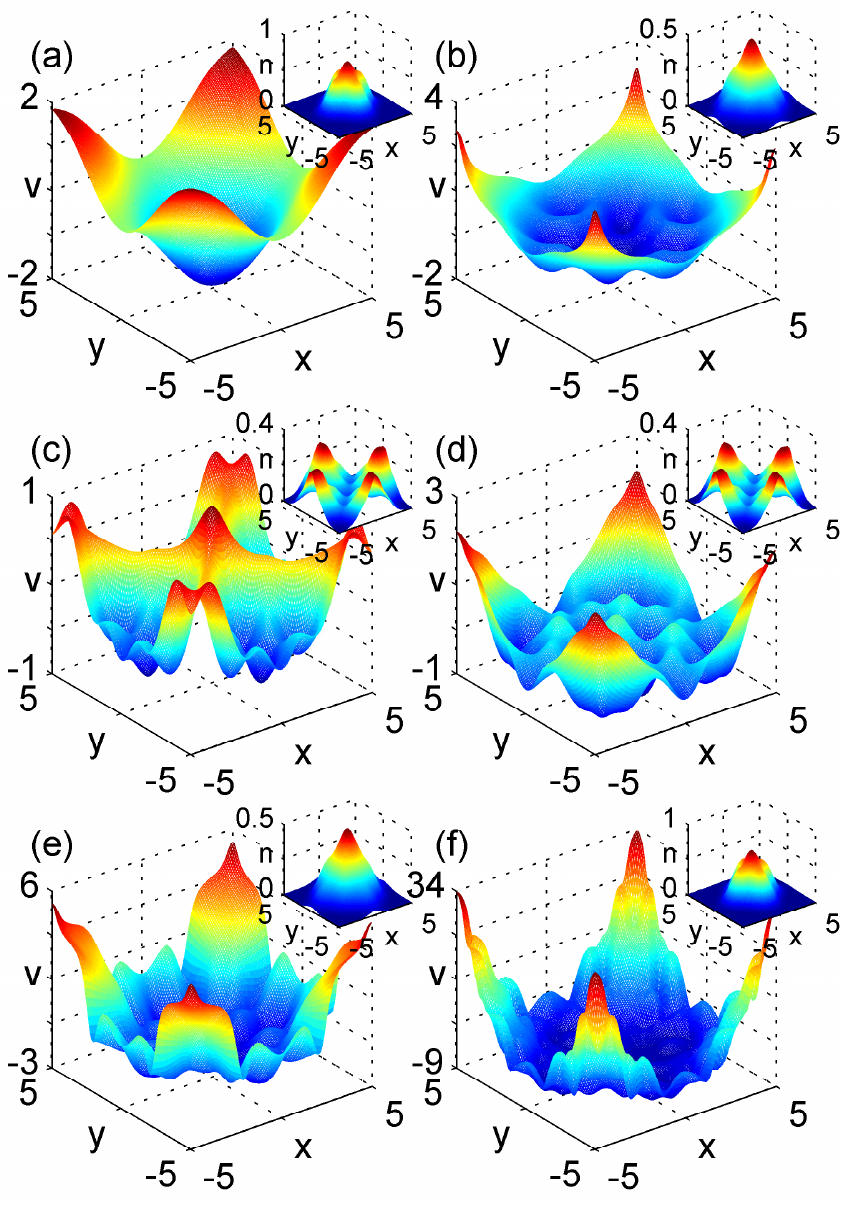}
\caption{(color online) Snapshots of the potential and splitted density (insets) for 10 non-interacting 2D particles at different times. (a) t=1,67, (b) t=5,
(c) t=8,33, (d) t=11,67, (e) t=15 and (f) t=18,33.}
\label{fig:NIPotentials2D}														
\end{figure}

Here we consider the 10 electron ground state of the above $v_0$ for $\lambda = 0$. The orbitals that form
this multi-particle state are easily found as they are the products of the orbitals of \cref{fig:orbitals}. 
This is the case, since $v_0$ decouples in an $x$ and $y$ component. The resulting external potential
that splits the corresponding ground-state density into four parts is illustrated in \cref{fig:NIPotentials2D}. Again
we point out, that many more snapshots would be required to give a representative presentation of the 2D
potential. 

%--------------------------------------------------------------------------------------------------------------------------------------------------%
\subsection{Optimal Control Theory}													
%--------------------------------------------------------------------------------------------------------------------------------------------------%

Here we will first give an example of our general approach to OCT, i.e., how we can restrict the search
space, and afterwards we will give a simple analytic example of how we can very easily treat the special case,
where the functional can be expressed by the density only, by first optimizing the density and then using 
our routine.

\subsubsection{By Restricted Search}

We consider again the interacting two-particle system of \cref{sec:Example1DModelInteracting}. Now, however, 
instead of prescribing the path $r(t)$ we want to find an optimal path that either translates the density around 
the ring in the given time $T$, or splits the density into equal halves and rejoins them at $T$, while minimizing the 
external field strength functional $J[n] = \int_{t_0}^{T } d t \int d x (\partial_x v([\Psi_0,n],x t))^2$. To do this
we employ the basis $r_n(t) = 0.5L(1.0-\cos((2n+1)\pi t/T)$. This way $r(t)$ becomes a function of 
the coefficients $c_n(t)$,
\begin{equation}
r(t) = \left( 1 -\sum_{n=1}^{N_B} c_n \right) r_0(t) + \sum_{n=1}^{N_B} c_n r_n(t) ,			
\end{equation}
for $n>0$. Since we then can compute for every $n[r] = n[\{c_n\}]$ with our algorithm
the corresponding $v[r] = v[n]$, $J[r] $ and hence $J[\{c_n\}]$,			
we can minimize this last functional with respect to $\{c_n\}$. This means minimizing a 
function of $N_B$ variables, where each function evaluation requires calculating a $v[\Psi_0,n]$ with our algorithm.
There exist many numerical methods for such minimization procedures. We have used the very simple downhill-simplex
 method, and even with this very slowly converging method, calculations for interacting 2-electron systems are possible 
on a normal computer. This is numerically much cheaper then standard OCT methods, where there is no simple way to restrict 
the problem to a small basis set. We can adopt the freedom of motion and also the number of basis functions to match the 
available computer power.

\begin{figure} [H]													
\includegraphics[width=8.6cm]{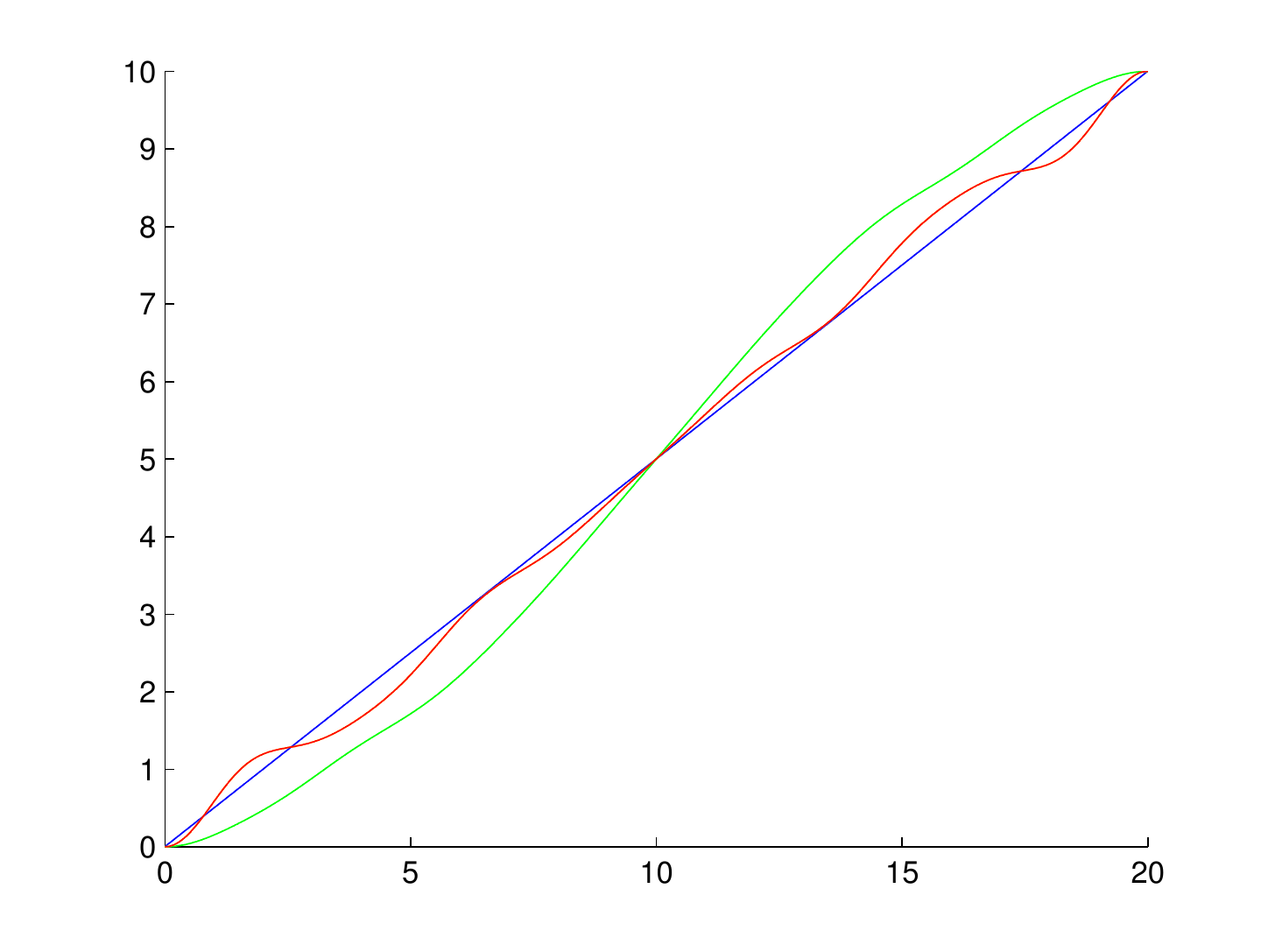}
\caption{(color online) The path $r(t)$ that minimize the external field energy for translation (green), splitting (red) and 
a linear path for reference (blue).}
\label{fig:OptimalPath}
\end{figure}

The optimal paths $r(t)$ are displayed in \cref{fig:OptimalPath}, where the green path corresponds to the optimal
translation and the red one to optimal splitting. For reference we have given the linear path in blue, which is actually the solution
of an analytic optimization presented in the next subsection.

\begin{figure} [H]
\includegraphics[width=8.6cm]{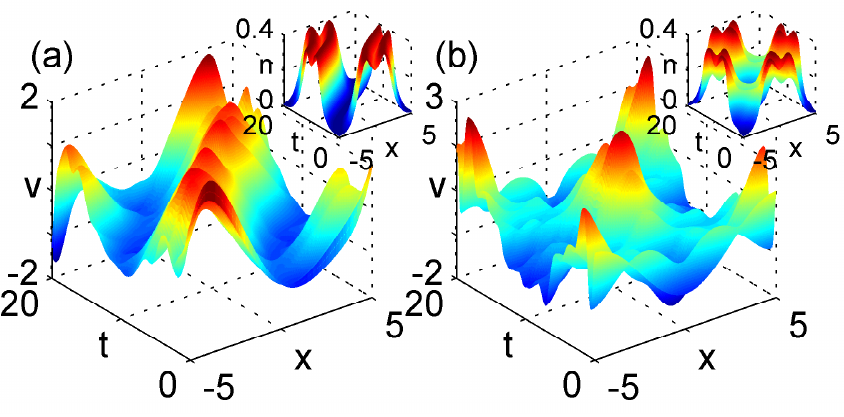}							
\caption{(color online) Optimal external potential for (a) translation and (b) splitting. Optimal densities are given as an inset.}
\label{fig:OptimalPotentials}
\end{figure}

In \cref{fig:OptimalPotentials} we have then displayed the external potentials and densities (in the inset) that correspond
to these optimal paths $r(t)$. 

\subsubsection{By Optimal Density}

Now we consider a case, where we can prescribe every constraint in terms of the density explicitly, and
thus can divide the problem in first determining an optimal density and then calculating the associated
potential with our algorithm.

We stay within the previous setting and ask for the optimal translated density $n(xt) = n_0(x-r(t))$.					
Then we only have to optimize $r(t)$, and the current is given by 
$j(xt) =  (\partial_t r(t)) n(xt)$. Now let us minimize  $\int_0^T \left( \int \partial_t j(xt) dx \right)^2 dt = N^2 \int_0^T \partial_t^2  r^2(t) dt$
(where we have used $\int n(xt)dx = N$ and $\int \partial_t n(xt)dx = 0$) 
subject to $r(0) = \partial_t r(0) = \partial_t r(T) = 0$ and $r(T) = L$. 
The analytic solution to this optimization problem is
\begin{equation*}																
r(t) =  - 2L\frac{\left( t - \tfrac{1}{2}T \right)^3}{T^3} + \frac{3L}{2}\frac{\left( t - \tfrac{1}{2}T \right)}{T} + \frac{L}{2}
\end{equation*}
and one can then use our density LCT to determine the corresponding potential. A further
such analytic minimization can be found in the case of the constraint 
$\int_0^T dt \int j^2(xt) dx = \int_0^T \partial_t r^2(t) dt$, which leads (although it violates the initial
and final conditions) to the linear solution $r = L t/T$, which is depicted also as a reference in \cref{fig:OptimalPath}.

While these problems are a bit artificial, it shows us that we can indeed split the OCT problem up into finding 
an optimal density and then use tracking to determine the potential, provided the functionals can be expressed in 
terms of the density.

%--------------------------------------------------------------------------------------------------------------------------------------------------%
\section{Conclusion \& Outlook}
%--------------------------------------------------------------------------------------------------------------------------------------------------%

In this work we have extensively discussed and generalized a recently 
introduced method \cite{Nielsen2012} to construct the external 
potential $v$ that, for a given initial state, produces a prescribed 
time-dependent density in an interacting quantum many-body system. 
We focused on its potential in control theory, and argued for its great 
efficiency and flexibility. Basically, the simple interpretation of the 
density (the amount of electrons per volume) allows us to use our
physical intuition to consider many interesting LCT phenomena, and 
to restrict the search space in OCT for efficiency. The close connection 
of this density-based control method with TDDFT makes studies of 
large systems efficient and realisable. We have further extended the 
method to 2D and 3D, and discussed the numerical implementation 
in great detail. We have also presented several examples to illustrate
the flexibility, and to confirm that the scheme is efficient and stable 
even for large and rapid density variations irrespective of the initial 
state and interactions.

While our present implementation of the algorithm focuses 
mainly on theory development and simple model systems, it can be 
easily adopted to also consider realistic systems. It seems possible that
this density-based control method can be used to predict control fields
for real systems. A possible application could be ultracold quantum gases, 
where the electric potentials to manipulate Bose-Einstein condensates 
can be well-controlled temporally and spatially.

\begin{acknowledgments}																		
S.E.B.N. and M.R. acknowledge financial support by the Austrian Science Fonds (FWF projects J 3016-N16 and P 25739-N27). 
R.v.L acknowledges the academy of Finland for support.
\end{acknowledgments}

\appendix

%---------------------------------------------------------------------------------------------------------------------------------------------------------------------------------%
\section{Boundary Conditions} \label{sec:BC}											
															%---------------------------------------------------------------------------------------------------------------------------------------------------------------------------------%

In this \namecref{sec:BC}, we will discuss how to properly and efficiently apply the formal time-step operator 
$\hat U(\Delta t) = e^{-i\hat{H}\Delta t}$, how to restrict the initial state and potential to ensure the 
resulting time-dependent wave function satisfies various basic physical properties,	
and how boundary conditions (BCs) are needed on a finite interval to ensure these properties. Surprisingly these
rather abstract considerations have real numerical consequences and if not taken into account
can lead to a breakdown of the algorithm.

Given the spectral form of a Hamiltonian,									
\begin{equation} \label{spectralHamiltonian}
\hat{H} = \sum_n E_n \ket|n>\bra<n| ,
\end{equation}															
where $\ket|n>\bra<n| $ is a projection onto the eigenstate $\ket|n>$,
the proper way to apply the time-step operator is to use,
\begin{equation} \label{TimeStepOperatorEigen}									
\hat U(\Delta t) = \sum_n e^{-iE_n \Delta t} \ket|n>\bra<n| .
\end{equation}
Thus we need to expand the state in the eigenstates $\ket|\Psi> = \sum_n \Psi_n \ket|n>$,
and use that the time-evolution of each eigenstate is given by a simple exponential, 
characterised by its energy $E_n$. This works for any normalisable state, 
$\langle \Psi \vert \Psi \rangle < \infty$, and correctly keeps the normalisation
per construction, if the eigenstates $\ket|n>$ form a basis for all normalisable $\ket|\Psi>$, and all the 
eigenvalues $E_n$ are real, i.e., $\hat{H}$ is self-adjoint. To ensure these properties in our case
of Hamiltonians of the form  \labelcref{Hamiltonian} we usually need to provide BCs due 
to the nature of the kinetic-energy operator as a differential operator\footnote{The case of an 
infinite volume is slightly different, since then the kinetic-energy operator has 
non-normalizable (which are also called distributional 
or scattering) eigenstates, i.e., the Laplacian then has a continuous spectrum. Since then also the 
spectral representation is slightly different we disregard this more subtle case.}.

For instance, consider a single free electron in one dimension on a finite interval $]0,L[$ . 
The differential operator $-\frac{1}{2} \partial_x^2$ has general eigenfunctions of the form 
$f_k(x) = a \exp(-i k x) + b \exp(i k x)$. By choosing two appropriate BCs (that fix $a$ and $b$ 
up to normalization) we single out a unique eigenbasis for all square-integrable wave functions
(otherwise we would have too many functions and the $f_k$ would not form a usual basis). 
While we can think of a lot of different BCs, we are in this work mainly concerned with 
either zero BCs, i.e., $f_k(0) = f_k(L) = 0$, which leads to the eigenbasis
\begin{eqnarray}
\langle x \vert k \rangle = \frac{2}{L} \sin\left(\frac{k \pi x}{L}  \right), \quad k \in \mathbb{N}
\end{eqnarray}
and the periodic BCs, i.e., $f_k(0) = f_k(L) $ and $\partial_x f_k(0) = \partial_x f_k(L)$,
which leads to the eigenbasis
\begin{eqnarray}
\langle x \vert k \rangle = \frac{1}{L} e^{i 2 \pi k x/L}, \quad k \in \mathbb{Z}.
\end{eqnarray}
Consequently the two different resulting self-adjoint Hamiltonians of the form 
$\hat H = \sum_k E_k  \ket|k>\bra<k|$ can then be used to propagate any normalizable
initial state (and for the same wave function lead to two different time-propagations).

Now, while \cref{TimeStepOperatorEigen} works for any normalisable $\ket|\Psi>$,
not all such states actually have finite energy, nor does the inner product 
$\bra<\Phi|\hat{H} \ket|\Psi>$ have sense for all such states, as it for example also can 
be infinite (e.g. when $E$ is)\footnote{As an example, we consider the case where $E_n=n^2$ and $n \in \mathbb{N}$,
corresponding to the above case of a free electron in 1D with zero BCs (and $L=\pi$).
While $\ket|\Psi> = \sum_n \frac{1}{n} \ket|n>$ is normalisable since 
$\langle \Psi \vert \Psi \rangle = \sum_n \frac{1}{n^2} < \infty$, it has infinite energy 
$\bra<\Psi|\hat{H} \ket|\Psi> = \sum_n 1 \rightarrow \infty$.
If we also introduce $\ket|\Phi> = \sum_n \frac{(-1)^n}{n} \ket|n>$, then 
$\bra<\Phi|\hat{H} \ket|\Psi> = \sum_n (-1)^n$, which does not converge, revealing 
another way the inner product can be meaningless.} Consequently, for such states also
$\langle \Phi \vert \hat{H} \Psi \rangle = \langle \hat{H}\Phi \vert \Psi \rangle$ no longer has sense in general.
This, however, is a basic requirement that our Hamiltonian is self-adjoint. These problems
arise when we apply the Hamiltonian outside of its domain $\mathbb{D}(\hat{H})$, i.e.,
the set of normalizable functions $\ket|\Psi>$ with $\hat{H} \ket|\Psi>$ normalisable again\footnote{If 
$\langle \hat{H}\Psi  \vert \hat{H} \Psi \rangle = \sum_n E_n^2 |\Psi_n|^2$ is finite for $\ket|\Psi>$ 
and $\ket|\Phi>$, and they are normalisable, then $\left| \bra<\Phi|\hat{H} \ket|\Psi> \right| \le \sum_n |E_n| |\Phi_n| |\Psi_n| 
\le \sum_n \left(1+E_n^2 \right) \left(|\Phi_n|^2 + |\Psi_n|^2 \right) < \infty$,
ensuring the inner product has sense and the energy is finite.}. Hence to ensure these, 
and many other, basic physical properties \footnote{Like the continuity equation and 
canonical commutation relations (between the position and momentum operator)}
a sufficient condition is $\ket|\Psi> \in \mathbb{D}(\hat{H})$.

For time-propagation with a Hamiltonian of the form \labelcref{Hamiltonian}
this means we must restrict the initial state $\ket|\Psi_0>$ and external potential $\hat{V}(t)$
to ensure that the wave function $\ket|\Psi>$ stays in $\mathbb{D}(\hat{H})$ at all times.
The easiest way to do this, is to ensure that $\hat{T} \ket|\Psi>$, $\hat{V} \ket|\Psi>$ and $\hat{W} \ket|\Psi>$ 
are individually normalisable\footnote{Besides the form of the kinetic-energy operator
we also usually cannot control the way the particles interact. Thus this condition
on $\hat W$ obviously restricts which interacting systems we can consider. On the other hand, 
since we can usually control the external potential $\hat{V}$, the above condition limits what we can do
to the system, e.g., potentials that are too singular might not be allowed.}.											
Hence, we limit $\hat{V}$ and $\hat{W}$ to those for which $\hat{V} \ket|\Psi>$ and $\hat{W} \ket|\Psi>$ 
are normalisable for all $\ket|\Psi> \in \mathbb{D}(\hat{T})$\footnote{In terms of the
eigenbasis $\ket|\bf k>$ of $\hat T$ the simple conditions for a function to be in the domain  are $\sum_\mathbf{k} |\Psi_\mathbf{k}|^2
 < \infty$ and $\sum_\mathbf{k} \mathbf{k}^2 |\Psi_\mathbf{k}|^2 < \infty$.}. Then $\mathbb{D}(\hat{H}) = \mathbb{D}(\hat{T})$, 
i.e., $\hat{H}\ket|\Psi>$ is normalisable if and only if $\hat{T}\ket|\Psi>$ is. By \cref{TimeStepOperatorEigen} it then
follows that $\hat{U}(\Delta t)\ket|\Psi> \in \mathbb{D}(\hat{T})$ if and only if 
$\ket|\Psi> \in \mathbb{D}(\hat{T})$\footnote{Since $\mathbb{D}(\hat{H})=\mathbb{D}(\hat{T})$, 
this follows from the fact that $\hat{H} \ket|\Psi>$ and $\hat{U}(\Delta t)\ket|\Psi>$ have the same norm,
i.e., $\langle \hat{H}\hat{U}(\Delta t)\Psi \vert \hat{H}\hat{U}(\Delta t)\Psi \rangle = \langle \hat{H}\Psi \vert \hat{H}\Psi \rangle$.}.
Using any of the time-stepping procedures of \cref{sec:TP}, which simply use multiple applications 
of \cref{TimeStepOperatorEigen}, with different $E_n$ and $\ket|n>$, it is then clear that if 
$\ket|\Psi_0> \in \mathbb{D}(\hat{T})$, then $\ket|\Psi> \in \mathbb{D}(\hat{T}) = \mathbb{D}(\hat{H})$ 
at all times, exactly as desired.
															
What is now left is to find simple conditions on the external potential and interaction such that
they have the same domain as the kinetic-energy operator. For this we need a characterization
of the domain $\mathbb{D}(\hat T)$ in terms of a spatial representation since the potentials and 
interactions are usually defined as a multiplicative operator in a spatial basis. We first take a 
look at our initial example of a particle on an interval $]0,L[$. There we have characterised the 
eigenfunctions by their respective BCs. As we discussed before, a sufficient condition for a
normalizable state $\ket|\Psi>$ to be in the domain of these kinetic-energy operators is that the 
eigenexpansion fulfils $\sum_k E_k^2 |\Psi_k|^2 < \infty$. Obviously
any state for which $|\Psi_k| = 0$ for all $k > K$, i.e., consisting of finitely many eigenfunctions, will
automatically fulfil this condition and be part of $\mathbb{D}(\hat T)$. And it is also obvious that
any such state will have the same BCs as the eigenfunctions. In fact, it can be shown \cite{blanchard2003mathematical} 
that all functions in the domain of the kinetic-energy operator obey the same BCs as the
eigenfunctions. Thus, any twice-differentiable wave function $\Psi(x)$ that obeys the BCs is
in the domain of $\mathbb{D}(\hat T)$, and we see by partial integration that the basic physical 
condition 
\begin{align}
\langle \Phi \vert \hat{T}\Psi \rangle &= \int_0^L dx \, \Phi(x)^*\left(-\frac{1}{2}\partial_x^2\right) \Psi(x)
\\ &=\int_0^L dx \, \left(\left(-\frac{1}{2}\partial_x^2\right)  \Phi(x)\right)^* \Psi(x)= \langle \hat{T}\Phi \vert \Psi \rangle \nonumber
\end{align}
is fulfilled for all functions in this set\footnote{We point out, that only for wave functions in the domain of the kinetic-energy operator
the equivalence $\sum_k E_k \langle x \vert k \rangle \Psi_k \equiv -\frac{1}{2} \partial_x^2 \Psi(x)$ holds. 
Outside of the domain the self-adjoint operator is not well-defined while $ -\frac{1}{2} \partial_x^2 \Psi(x)$ 
might still exist.}. Thus while the basic operation $-\frac{1}{2} \partial_x^2$ is the same for any kinetic-energy
operator, e.g., either zero or periodic BCs, the set of functions for which it is self-adjoint can be different. Thus
if we want to stabilise the domain during time-propagation, this means we need to keep the chosen boundary 
conditions for all times.

While we argued now in terms of a one-dimensional problem, we point out that the eigenfunctions of the
kinetic-energy operator in higher dimensions are just products of the one-dimensional eigenfunctions. Hence
the previous discussion equally well applies to a multi-dimensional box $]0,L[^{3N}$. Further the previous consideration
directly leads to another way of seeing the BCs, namely, instead of enforcing a certain behaviour of the wave 
functions on the boundary, the solutions need to obey a certain periodicity in the whole space. This is obvious for the 
periodic BCs, which imply that the wave functions have to be repeated periodically on the whole space,
while (maybe less known) for the zero BCs, we see from the behaviour of $\sin(k \pi x /L)$ that the solutions need to 
be periodic on $2L$, and odd about all endpoints of an subinterval with length $L$.

With this characterization of the physical wave functions in terms of differentiability and BCs we can easily consider
conditions on the potentials $v(\br t)$ and $w(|\br|)$ to stabilize the domain of the kinetic-energy
operator. A simple (but not the most general) condition is that $v$ and $w$ are bounded functions. Employing, 
however, the fact that the wave functions are twice differentiable one can show \cite{blanchard2003mathematical} that square-integrability
of the potentials and interactions is enough. It is important to note, that mathematically no specific differentiability 
or boundary conditions have to be imposed on the potentials and interactions. However, if we view the finite-interval
problem extended to the whole space, we see that the potential and the interaction have to be extended
periodically in the periodic BCs case and even and double-periodic in the zero BCs case.

So this result is great, it is both well-known and very general.		
There is but one issue. Namely that any computer would get a heart-attack, 
if we asked it to compute all the eigenstates for anything but the simplest 
systems. And it does not exactly help that we have to repeatedly recalculate them
(as the Hamiltonian differs in each step since $v$ changes).
So numerically, we usually want to use a simpler form of the 
evolution operator in spatial representation,					
and restrict ourselves to analytic potentials and initial states\footnote{This
is not a real restriction since we can approximate any square-integrable function
by analytic functions to any accuracy we want, e.g., one can use 
the eigenfunctions of the kinetic-energy operator.}.						
This allows us to formally use the much more effective 
time-step operator
\begin{equation} \label{TimeStepOperatorAnalytic}												
\hat U(\Delta t) = \sum_n \frac{(-i \hat H \Delta t)^n}{n!} ,
\end{equation}
where we apply $\hat H$ by directly using the right-hand side of \cref{Hamiltonian} instead 
of by using its eigen-expansion as in \cref{TimeStepOperatorEigen}\footnote{Note 
that if $\hat H$ is interpreted by its eigen-expansion in \cref{TimeStepOperatorAnalytic},
it is still only equivalent with \cref{TimeStepOperatorEigen} if one can exchange the order of summation,			
which cannot always be done, even if the state the Hamiltonian acts on is nicely complex analytic \cite{Kowalevskaya}.						
Further, the reason \cref{TimeStepOperatorAnalytic} does not apply to a general state in the domain
of the Hamiltonian is that while acting once with $\hat H$ on the state keeps it normalised, multiple
applications of the Hamiltonian to the very same state can make it non-normalizable.}. In this case, 
we must enforce the BCs by further restricting the initial state, external potential as well as the 
interaction to ensure the the wave function stays analytic and keeps satisfying the desired BCs during 
time-stepping, i.e., after applying \cref{TimeStepOperatorAnalytic}. Thus if we want the wave function 
to satisfy periodic BCs, the initial state and potential must fulfil periodic BCs. \footnote{Since the initial 
state and potential are analytic, periodic BCs now means that all derivatives must be the same at $0$ and $L$.} 
If we instead want the wave function to satisfy zero BCs, things get more complicated.					
To better understand this, let us first consider the case of free propagation ($v=0$),
and limit us to a single electron in 1D, since the generalisation to more electrons and higher dimensions 
is straightforward.	The Taylor expansion of the initial state at the boundary $x=0$ must then not contain 
any even term $x^{2k}$, since applying \labelcref{TimeStepOperatorAnalytic} leads to terms 
of the form $x^{2k-2n}$, which could then be non-zero at the boundary.
Such terms would then lead to a non-zero wave function at the boundary
and thus violating the zero BCs. This is also true when $v \ne 0$, since if we have a $x^2$ 
component in $\Psi$ at $x=0$, the only potential that can make $\hat H \Psi$ stay zero at the boundary 
is of the form $v\sim x^{-2}$, which is not allowed by our basic mathematical considerations.				
In this manner we can rule out all higher even components of the wave functions. 
The odd terms of the wave functions are not problematic since they can never lead to 
a non-zero terms at $x=0$. Hence, to ensure that we have an odd wave function about the boundary
the potential $v$ must have no $x^{2k+1}$ components. In this case the multiplication with the external
potential never leads to even components of the wave functions at $x=0$.										
To conclude, $\Psi$ must always be odd over the boundaries while the potential $v$ has to be 
even\footnote{Using sufficiently many even terms, one can approximate odd potentials arbitrarily well.
So analytically one can also treat potentials with odd components, and numerically one can get 
very close to these results.}.

It is curious how these numerical Even-Odd restrictions are not mentioned in the general literature.
However, given that a wrong treatment of the boundaries only will lead to errors in the (usually) irrelevant 
region near the boundary where $\Psi \sim 0$, it hardly matters if one is doing time-propagation 
(even if $\Psi$ is wrong by orders of magnitude in this region). However, when doing our density-based
LCT scheme, it is crucial to also treat the low density region correctly.	
Otherwise the potential can go completely crazy at the boundaries, since to get the correct wave function 
while violating the BCs requires the potential to compensate the errors and can lead to artificially large potentials.
Since the relative errors that must be compensated tend to be large, this happens although the wave function
is small near the boundary. These large and artificial potentials extremely quickly 
influence the relevant region far away from the boundary, leading to a complete breakdown 
of the procedure.

It is nice to see how the BCs also play a major part in ensuring that the inversion has sense.
To invert the Sturm-Liouville (SL) equation (which together with time-propagation 
is actually all we need to do our density-based LCT method),
\begin{equation}											
- \nabla \cdot \left( n(\br t) \nabla v(\br t) \right) = \zeta(\br t) ,			
\end{equation}
where $\zeta(\br t)$ represent any of the right-hand sides of \cref{fp,fpn,fpnsimple,fpjsimple,fpnjsimple,fpnjsimplenum},
we must impose BCs that are consistent with basic physical requirements. By this we mean, that we need to impose BCs 
that are consistent with the fact that a change of gauge, i.e., adding a constant $c(t)$ to the potential, does not matter 
as well as that the resulting potential allows to perform the propagation without violating any constraints, e.g., the BCs. 
The first condition is required to be able to invert the problem uniquely. Since the SL operator
has zero eigenvalues we can only invert the problem perpendicular to the zero eigenfunction. Further,  
the inhomogeneities $\zeta(\br t)$ integrate to zero and are thus perpendicular to $c(t)$. We therefore want
to choose BCs that make $c(t)$ the unique zero eigenfunction. That this is possible in our case can be most easily
demonstrated in the 1D case, where the most general zero eigenfunction reads as										
\begin{equation}		
v_\mathrm{Hom}(xt) = c(t) + d(t) \int_a^x \frac{dy}{n(yt)}. 
\end{equation}	
To see that $d(t)$ vanishes for zero and periodic BCs, implying the correct freedom,
simply note $\int_a^b \frac{dy}{n(yt)} > 0$ as $n(yt) \ge 0$, so $v_\mathrm{Hom}$ 
cannot be periodic unless $d(t)=0$. Since we want to have periodic potentials (obviously
for the case of periodic wave functions, but also in the case of zero-BCs wave functions
where we are periodic on twice the interval) we accordingly have 
$v_\mathrm{Hom}(xt) = c(t)$. Further, for any square-integrable inhomogeneity we can 
invert the above SL equation perpendicular to the zero eigenfunction \cite{penz2011domains,GFPP1D}. 
The resulting potential by construction obeys then the appropriate BCs. Thus the 
inversion fits with the conditions on our propagation.

%--------------------------------------------------------------------------------------------------------------------------------------------------%
\section{Numerical Spatial Representation} \label{sec:SR}
%--------------------------------------------------------------------------------------------------------------------------------------------------%

Let us first consider periodic BCs since these are the easiest to implement.
Thus, if we for simplicity first consider the case of one electron in one-dimensional space, i.e. on a ring,			
an obvious way to represent the wave function numerically is to introduce an equidistant grid on the ring,
and then represent the wave function by its values at these points.
Now, why do we prefer this particular spatial representation?
Well, we want the wave function to be able to travel all the way around the ring, and be equally well 
represented wherever it is, which is exactly what this representation achieves.
Further, in order to do the time-stepping, we really only need to be able to perform one operation 
on the wave function, namely apply the Hamiltonian to the wave function. But this, and many other operations, 
are easily performed in this representation, since to apply the potential operator to the wave function we 
simply multiply the value of the potential and wave function at the given point.
To apply the kinetic operator, we need to be able to take (second order) spatial derivatives, but this is also 
easily done in the given representation, e.g. by the use of finite differences (usually we employ a 7th oder finite differencing scheme) 
or fast Fourier transforms and multiplication in momentum space. Finally it is obvious how one should incorporate 
the BCs into the numerical definition of the Hamiltonian in this representation,
since all we have to do is, that when we want to take a finite difference at a point near the boundary, we use points on both sides
(using fast Fourier transforms it is incorporated naturally into the transforms).
So although other better spatial representation may exist for specific cases, this choice is really the 
obvious general purpose spatial representation. Clearly, for one electron in 2D, we simply use a 2D grid.
Likewise, for two electrons in 1D, we also use a 2D grid, since we need one dimension for each spatial coordinate.
Finally, for two electrons in 2D, we need a 4D grid, and we are already reaching the limit of what a normal computer 
can handle, which is the price for being as general as we are.

%--------------------------------------------------------------------------------------------------------------------------------------------------%
\section{Time-Step Operators} \label{sec:TSO} % Approximate
%--------------------------------------------------------------------------------------------------------------------------------------------------%

Having picked a time-stepping strategy as described in \cref{sec:TP}, all that is left to do time-propagation is to approximate the time-step operator (exponential).
To obtain a good approximation, one should obviously try to build as much physics as possible into the approximation.		
Thus it is only natural to require that the approximation, at least to a high extend, should conserve the norm, meaning it must 
be unitary\footnote{A norm-conserving scheme is also often said to be neutrally stable (a stable scheme being one that conserve or decrease the norm).}.
In practise, this more or less only leaves a few options, here presented in order of strongly increasing efficiency 
(quite naturally coinciding with the chronological order, all these having been the method of choice at a time)
\footnote{A different scheme, which does not fit into the given structure, deserves mentioning.
This scheme, based on Chebyshev polynomial expansion, is strongly recommended for time-independent Hamiltonians,
but not for time-dependent Hamiltonians, so it is of no use to us.							
\[ \hat U(t) = e^{ - i \hat H t} \approx \sum\limits_{n = 0}^N a_n P_n( - i \hat H t) \]}.
We strongly recommend the use of the Lanzcos method, since its flexibility, superior efficiency and complete generality is well worth 
the slightly more involved implementation. The split-operator method then forms a very nice backup for controlling
the time-stepping.  We discourage the use of the Crank-Nicholson method, due to its poor efficiency and limitation to one electron in 1D,			
and the Second-Order Differencing method, due to its lack of true norm-conservation, and corresponding potential for oscillations.	
Indeed, these two methods are mainly included here to point out their limitations.

\begin{enumerate}
\item	
Crank-Nicholson (CN):\\ 
In this case we approximate the exponential by
\begin{equation*}
\frac{1 - \tfrac{i}{2} \hat{H}_{+1/2} \Delta t}{1 + \tfrac{i}{2} \hat{H}_{+1/2} \Delta t} ,
\end{equation*}
which is unitary as desired. To apply this operator, we have to solve a linear implicit equation for $\Psi_{+1}$,	
\begin{equation*}
\left( 1 + \tfrac{i}{2} \hat{H}_{+1/2} \Delta t \right) \Psi_{+1} = \left( 1 - \tfrac{i}{2} \hat H_{+1/2} \Delta t \right) \Psi .
\end{equation*}
This is very involved compared with an explicit equation, except if $\hat H$ is very sparse, and we can take advantage of this.
Originally, this method was thus designed for one electron in 1D, using a 3 point finite difference for $\hat{T}$,	   
resulting in a tridiagonal matrix representation of the Hamiltonian, so it can be solved almost as efficiently as an explicit equation.
While this can be extended to higher dimensions and more electrons using the alternating direction method, it is very involved. 
The method is thus essentially limited to one electron in 1D, and in this case present computers can solve the implicit equation 
very fast (even if the matrix is not  tridiagonal).

\item
Second-Order Differencing (SOD):\\ 
Using strategy C in \cref{sec:TP} and approximating the time-step operator by a first order Taylor expansion we find
\begin{equation*}
\Psi_{+1} = \Psi_{-1} - 2i \Delta t \hat H \Psi .
\end{equation*}
The only difference is that we  here use $\Psi$ instead of $\Psi_{-1}$ in the last term.							
This makes the expression exact to second order, since it may also be obtained using a central 3 point finite-difference formula 
for the time-derivative in the TDSE.													
Besides restricting us to strategy C, and hence to an equidistant grid, this method is also not norm-conserving.	
However, we may also write this method as,											
\begin{equation*}
\begin{pmatrix}
\Psi_{+1} \\
\Psi_{+2}
\end{pmatrix}
=
\begin{pmatrix}
1				& -2i \Delta t \hat{H_0} \\										
-2i \Delta t \hat{H_1}	& 1 - 4 \Delta t^2 \hat{H}_1 \hat{H}_0
\end{pmatrix}
\begin{pmatrix}
\Psi_{-1} \\
\Psi_0
\end{pmatrix} ,
\end{equation*}
where this operator is conditionally unitary. Thus, the oscillation of strategy C are especially bad here.		
This method was popular, as it (in contrast to CN) is explicit and can be used completely generally.		
Straight away it is also very simple to implement. However, the oscillations are often a serious issue,
and trying to improve this aspect, e.g. by using $<\Psi_{-1}|\hat{O}|\Psi_0>$ to calculate expectation values (including the norm),
it quickly becomes complicated. Higher order versions exist, but are more involved, and usually perform the same.

\item
Split-Operator (SPO):\\ 
Here we split the operator into the clearly unitary
\begin{equation*}
e^{ - \tfrac{i}{2} \hat V_{+1/2} \Delta t} e^{ - i \hat K \Delta t} e^{ - \tfrac{i}{2} \hat V_{+1/2} \Delta t} .
\end{equation*}
Assuming we use an equidistant-grid representation,
to apply this operator we first apply the first part by multiplication.
We then do a Fourier transform (periodic BCs) or sine transform (zero BCs),
and then apply the kinetic part by multiplication in Fourier space, 
whereafter we transform back, and apply the last potential by multiplication
on the real-space grid again. This method is simply and effective, however, 
the Lanczos method is still usually better.

\item

Lanczos:\\ 
This method is special since it approximates $e^{-i\hat{H}\Delta t} \Psi$, which is what we really want 
to compute anyway, instead of just $\hat{H}$. This leads to higher precision, since it makes use of more 
information (namely $\Psi$)\footnote{Unfortunately it also breaks the time-reversal symmetry slightly,
as the approximation will then depend on the non-midpoint $\ket|\Psi>$,								
and hence differ going back and forth in time. But it is a minor problem since
 the approximation is very good and only depends slightly on $\ket|\Psi>$. 
Further$\ket|\Psi>$ changes little in a time step.}.		
The idea is to first construct an orthonormal basis $q_i$ for the so-called 
$N$-dimensional Krylov subspace, i.e., for the subspace spanned by $\hat H_i \Psi$, for $i \in [0,N]$
(the optimal value of $N$ depends on the problem, but typically lies between say 6 and 20).					
To do this, start with $q_0=\Psi$, and for $i \in [1,N]$ compute $\hat{H} q_{i-1}$, and make it orthonormal 
to all $q_j$, for $j \in [0,i-1]$\footnote{Note $\bra<q_i|\hat{H} \ket|q_j>$ is tri-diagonal,
since $\hat H$ is Hermitian and $\bra<q_i|\hat{H} \ket|q_j>=0$ for $i \ge j+2$, due to the fact that $\hat H \ket|q_j>$ 
only contains $\ket|q_k>$ with $k \le j+1$. So to construct $\ket|q_i>$, we actually only need 
to make $\hat{H}\ket|q_{i-1}>$ orthonormal to $\ket|q_{i-1}>$, since it is already orthogonal to all $\ket|q_j>$ for $j \le i-2$.}.

Now the approximation is to employ $\sum_i \ket|q_i>\bra<q_i| e^{-i\hat{H} \Delta t} \ket|q_0>$
to finite order, i.e., the true result is approximated by its projection onto the Krylov subspace.
But this is a very good approximation, since in the short time $\Delta t$,
very little of the wave function is likely to leave the Krylov subspace.
\footnote{Unfortunately, what does leave the subspace courses a norm loss,
so the approximation is not strictly norm-conserving, only stable. However, the loss is in principle a 
direct measure of the precision, so it is clear when there is a problem, and it is usually so small,								
that it gets washed out by other effects.}
 
To compute the row $\bra<q_i| e^{ - i \hat H \Delta t} \ket|q_0>$ of the submatrix $\bra<q_i| e^{ - i \hat H \Delta t} \ket|q_j>$,
note that $\bra<q_i| e^{ - i \hat H \Delta t} \ket|q_j> = (V \exp( - i D_N \Delta t ) V^\dagger)_{ij}$,	
where we have introduced $D$ and $V$ as the diagonalisation of $\bra<q_i|\hat{H} \ket|q_j>$,			
i.e. $\bra<q_i|\hat{H} \ket|q_j> = (VDV^\dagger)_{ij}$.								

Lanczos is also especially easy and effective in connection with adaptive time steps,  since there is a simple, decent error estimate.

\end{enumerate}													
To compare the efficiency of these schemes, we first note that CN, SOD and SPO are all 
precise to second order. However, while CN and SOD do not care much about the higher 
order terms, apart from the requirement of norm-conservation,
SPO gives a decent approximation to these, allowing for significantly longer time-steps. 	
Lanczos performs even better since it is precise to the order of the Krylov subspace,							
and gets the part of the higher-order terms lying inside the Krylov subspace,
which is a very good approximation. So for fast changing external potentials, it is not uncommon 
that the limitation on $\Delta t$ is no longer by energies, but instead by the changes of $V(t)$ 
that must be resolved. To conclude, limits on $\Delta t$ for the different schemes are typically 
$\Delta t < \tfrac{1}{5} | E_{Grid} |_{\max}^{-1}$ (Eigenvalue) for CN and SOD,	for split-operator 
$\Delta t < | E_{Interest} |_{\max}^{-1}$ and for Lanzcos  we need to able to resolve $V(t)$ 
and $\Delta t < | E_{Krylov} |_{\max}^{-1}$

%---------------------------------------------------------------------------------------------------------------------------------------------------------------------------------%

\clearpage
\bibliographystyle{aip}
\bibliography{Library}

\end{document}